\documentclass[journal,twoside]{IEEEtran}
\usepackage{cite}
\usepackage{url}
\usepackage{graphicx} 
\usepackage{amsmath,amssymb} % define this before the line numbering.
\usepackage{algorithm} 
\usepackage{algorithmic}
\usepackage[normalem]{ulem}
\usepackage{multirow}
\usepackage{float}
\usepackage{overpic}
\usepackage{rotating}
\usepackage[table]{xcolor}
\usepackage[pagebackref=false,breaklinks=true,colorlinks, bookmarks=false]{hyperref}

\usepackage{pgfplots}
\pgfdeclarelayer{background}
\pgfdeclarelayer{foreground}
\pgfsetlayers{background,main,foreground}
\newlength{\iwidth}
\newlength{\iheight}
\newlength{\rowspacing}

\usepackage{marginnote} 

\newcommand{\tabincell}[2]{\begin{tabular}{@{}#1@{}}#2\end{tabular}}

\usepackage{silence}
\def\ie{\emph{i.e.}}
\def\eg{\emph{e.g.}}
\def\etc{\emph{etc}}
\def\etal{\textit{et al.}}
 
\definecolor{bblue}{rgb}{0,150,230}
\definecolor{mygray}{gray}{.92}

\newcommand{\tabref}[1]{Table~\ref{#1}}

\newcommand{\secref}[1]{$\S$ \ref{#1}}

\newcommand{\tr}[1]{\textbf{\textcolor{red}{#1}}}

\newcommand{\tb}[1]{\textcolor{blue}{#1}}

\def\ourdataset{\textit{COVID-SemiSeg}}
\def\ourmodel{\textit{Inf-Net}}
\graphicspath{{./Imgs/}{../../figure/}}
 
\begin{document}

\title{Inf-Net: Automatic COVID-19 Lung Infection Segmentation from CT Images}

\author{
Deng-Ping~Fan, 
Tao~Zhou, 
Ge-Peng~Ji, 
Yi~Zhou, 
Geng~Chen, 
Huazhu~Fu, 
Jianbing~Shen,  
and Ling~Shao%
\thanks{Corresponding authors: \textit{Geng Chen, Huazhu Fu, and Jianbing Shen}.}
\thanks{D.-P.~Fan, T.~Zhou, Y.~Zhou, G.~Chen, H.~Fu, and J.~Shen are with the Inception Institute of Artificial Intelligence, Abu Dhabi, UAE. (e-mails: \{dengping.fan, tao.zhou, yi.zhou, geng.chen, huazhu.fu, jianbing.shen\}@inceptioniai.org)}
\thanks{G.-P.~Ji is with School of Computer Science, Wuhan University, Hubei, China. (e-mail: gepengai.ji@gmail.com)}
\thanks{L.~Shao is with the Mohamed bin Zayed University of Artificial Intelligence, Abu Dhabi, UAE, and also with the Inception Institute of Artificial Intelligence, Abu Dhabi, UAE. (e-mails: ling.shao@inceptioniai.org)}
}

\maketitle
 
\begin{abstract}

Coronavirus Disease 2019 (COVID-19) spread globally in early 2020, causing the world to face an existential health crisis. Automated detection of lung infections from computed tomography (CT) images offers a great potential to augment the traditional healthcare strategy for tackling COVID-19. However, segmenting infected regions from CT slices faces several challenges, including high variation in infection characteristics, and low intensity contrast between infections and normal tissues. Further, collecting a large amount of data is impractical within a short time period, inhibiting the training of a deep model. To address these challenges, a novel COVID-19 Lung Infection Segmentation Deep Network (\ourmodel) is proposed to automatically identify infected regions from chest CT slices. In our \ourmodel, a parallel partial decoder is used to aggregate the high-level features and generate a global map. Then, the implicit reverse attention and explicit edge-attention are utilized to model the boundaries and enhance the  representations. Moreover, to alleviate the shortage of labeled data, we present a semi-supervised segmentation framework based on a randomly selected propagation strategy, which only requires a few labeled images and leverages primarily unlabeled data. Our semi-supervised framework can improve the learning ability and achieve a higher performance. Extensive experiments on our \ourdataset~and real CT volumes demonstrate that the proposed \ourmodel~outperforms most cutting-edge segmentation models and advances the state-of-the-art performance. 

\end{abstract}

%\begin{IEEEkeywords}
%COVID-19, CT image, infection segmentation, semi-supervised learning.
%\end{IEEEkeywords}

\section{Introduction}\label{sec:introduction}
\IEEEPARstart{S}{ince} December~2019, the world has been facing a global health crisis: the pandemic of a novel Coronavirus Disease (COVID-19)~\cite{Wang2020_Lancet,Huang2020}.  {According to the global case count from the Center for Systems Science and Engineering (CSSE) at Johns Hopkins University (JHU)~\cite{JHU_number} (updated 1 May, 2020), 3,257,660 identified cases of COVID-19 have been reported so far, including 233,416 deaths and impacting more than 187 countries/regions.} For COVID-19 screening, the reverse-transcription polymerase chain reaction (RT-PCR) has been considered the gold standard. However, the shortage of equipment and strict requirements for testing environments limit the rapid and accurate screening of suspected subjects. Further, RT-PCR testing is also reported to suffer from high false negative rates~\cite{Ai2020}. As an important complement to RT-PCR tests, the radiological imaging techniques, \eg, X-rays and computed tomography (CT), have also demonstrated effectiveness in both current diagnosis, including follow-up assessment and evaluation of disease evolution~\cite{Rubin2020,shi2020review}. Moreover, a clinical study with 1014 patients in Wuhan China, has shown that chest CT analysis can achieve 0.97 of sensitivity, 0.25 of specificity, and 0.68 of accuracy for the detection of COVID-19, with RT-PCR results for reference~\cite{Ai2020}.  Similar observations were also reported in other studies~\cite{2020200432,2020200034}, suggesting that radiological imaging may be helpful in supporting early screening of COVID-19.

%\begin{figure}[!t]
%    \center
%    \includegraphics[width =1\linewidth ]{Infection} 
%    \vspace{-15pt}
%    \caption{Example of COVID-19 infected regions in CT axial slice, %where the red and green regions denote the GGO, and consolidation, %respectively. The images are collected from~\cite{data_seg}.}
%    \label{img-cover}
%\end{figure}

\begin{figure}[t!]
	\centering
    \small
	\begin{overpic}[width=\columnwidth]{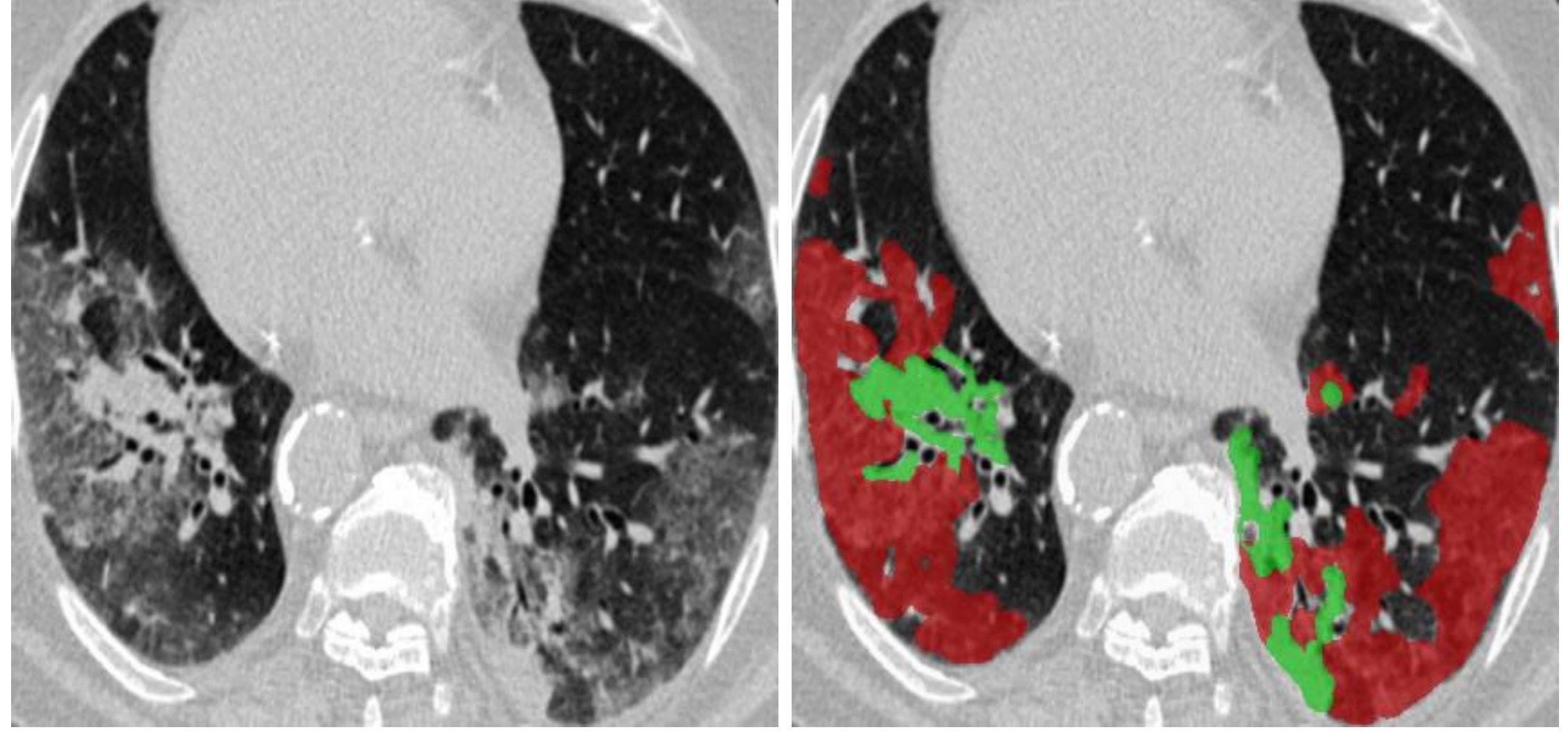}
    \put(22,-2) {(A)}
    \put(72,-2) {(B)}
    \end{overpic}
	\caption{Example of COVID-19 infected regions (B) in CT axial slice (A), where the red and green masks denote the GGO and consolidation, respectively. The images are collected from~\cite{data_seg}.}
    \label{img-cover}
\end{figure}

Compared to X-rays, CT screening is widely preferred due to its merit and three-dimensional view of the lung. In recent studies~\cite{Ai2020,Ye2020}, the typical signs of infection could be observed from CT slices, \eg, ground-glass opacity (GGO) in the early stage, and pulmonary consolidation in the late stage, as shown in Fig.~\ref{img-cover}.  The qualitative evaluation of infection and longitudinal changes in CT  slices could thus provide useful and important information in fighting against COVID-19. However, the manual delineation of lung infections is tedious and time-consuming work. In addition, infection annotation by radiologists is a highly subjective task, often influenced by individual bias and clinical experiences.

\begin{table}[t!]
  \centering
  \footnotesize
  \renewcommand{\arraystretch}{1.0}
  \setlength\tabcolsep{0.5pt}
    \caption{A summary of public COVID-19 imaging datasets. \#Cov and \#Non-COV denote the numbers of COVID-19 and Non-COVID-19 cases. $\dagger$ denotes the number is from~\cite{cohen2020covid}.
    }\label{tab:DatasetSummary}
  \begin{tabular}{l|c|c|c}
    \rowcolor{mygray}
  	\hline
  	Dataset                                          & Modality  & \#Cov/\#Non-COV &     Task     \\ \hline\hline
  	COVID-19 X-ray Collection~\cite{cohen2020covid}  &  X-rays   &      229$^\dagger$ / 0      &  Diagnosis   \\
  	COVID-19 CT Collection~\cite{cohen2020covid}     & CT volume &      20 / 0       &  Diagnosis   \\
  	COVID-CT-Dataset~\cite{zhao2020COVIDCT}          & CT image  &    288 / 1000     &  Diagnosis   \\
  	COVID-19 Patients Lungs~\cite{data_Patients}     &  X-rays   &      70 / 28      &  Diagnosis   \\
  	COVID-19 Radiography~\cite{RADIOGRAPHY_database} &  X-rays   &    219 / 2,686     &  Diagnosis   \\
  	COVID-19 CT Segmentation~\cite{data_seg}         & CT image  &      110 / 0      & Segmentation \\ \hline
  \end{tabular}
\end{table}

Recently, deep learning systems have been proposed to detect patients infected with COVID-19 via radiological imaging~\cite{shi2020review,bullock2020mapping}. For example, a COVID-Net was proposed to detect COVID-19 cases from chest radiography images~\cite{COVID-Net}.  An anomaly detection model was designed to assist radiologists in analyzing the vast amounts of chest X-ray images~\cite{Zhang2020_Anomaly}. For CT imaging, a location-attention oriented model was employed in~\cite{xu2020deep} to calculate the infection probability of COVID-19. A weakly-supervised deep learning-based software system was developed in~\cite{Zheng2020_weak} using 3D CT volumes to detect COVID-19. A paper list for COVID19 imaging-based AI works could be found in~\cite{COVID19-AI-Collection}.  Although plenty of AI systems have been proposed to provide assistance in diagnosing COVID-19 in clinical practice, there are only a few works related  infection segmentation in CT slices~\cite{Chaganti2020,shan+2020lung}. COVID-19 infection detection in CT slices is still a challenging task, for several issues: \textit{1) The high variation in texture, size and position of infections in CT slices is challenging for detection.} For example, consolidations are tiny/small, which easily results in the false-negative detection from a whole CT slices. \textit{2) The inter-class variance is small.} For example, GGO boundaries often have low contrast and blurred appearances, making them difficult to identify. \textit{3) Due to the emergency of COVID-19, it is difficult to collect sufficient labeled data within a short time for training deep model.} Further, acquiring high-quality pixel-level annotation of lung infections in CT slices is expensive and time-consuming. Table~\ref{tab:DatasetSummary} reports a list of the public COVID-19 imaging datasets, most of which focus on diagnosis, with only one dataset providing segmentation labels.

To address above issues, we propose a novel COVID-19 Lung Infection Segmentation Deep Network (\ourmodel) for CT slices.  Our motivation stems from the fact that, during lung infection detection, clinicians first roughly locate an infected region and then accurately extract its contour according to the local appearances. We therefore argue that the area and boundary are two key characteristics that distinguish normal tissues and infection. Thus, our \ourmodel~first predicts the coarse areas and then \textit{implicitly} models the boundaries by means of reverse attention and edge constraint guidance to \textit{explicitly} enhance the boundary identification. Moreover, to alleviate the shortage of labeled data, we also provide a semi-supervised segmentation system, which only requires a few labeled COVID-19 infection images and then enables the model to leverage unlabeled data. Specifically, our semi-supervised system utilizes a randomly selected propagation of unlabeled data to improve the learning capability and obtain a higher performance than some cutting edge models. 
In a nutshell, our contributions in this paper are threefold:
\begin{itemize}

\vspace{8pt}
\item We present a novel COVID-19 Lung Infection Segmentation Deep Network (\ourmodel) for CT slices. By aggregating features from high-level layers using a parallel partial decoder (PPD), the combined feature takes contextual information and generates a global map as the initial guidance areas for the subsequent steps. To further mine the boundary cues, we leverage a set of implicitly recurrent reverse attention (RA) modules and explicit edge-attention guidance to establish the relationship between areas and boundary cues. 

\vspace{8pt}
\item A semi-supervised segmentation system for COVID-19 infection segmentation is introduced to alleviate the shortage of labeled data. Based on a randomly selected propagation, our semi-supervised system has better learning ability (see \secref{sec:Experiments}).

\vspace{8pt}
\item We also build a semi-supervised COVID-19 infection segmentation (\ourdataset) dataset, with 100 labeled CT slices from the COVID-19 CT Segmentation dataset~\cite{data_seg} and 1600 unlabeled images from the COVID-19 CT Collection dataset~\cite{cohen2020covid}. Extensive experiments on this dataset demonstrate that the proposed \ourmodel~and \textit{Semi-}\ourmodel~outperform most cutting-edge segmentation models and advances the state-of-the-art performance. Our code and dataset have been released at: \href{https://github.com/DengPingFan/Inf-Net}{https://github.com/DengPingFan/Inf-Net}
\end{itemize}

\section{Related works} 

In this section, we discuss three types of works that are most related to our work, including: segmentation in chest CT, semi-supervised learning,  and artificial intelligence for COVID-19.

\subsection{Segmentation in Chest CT}
CT imaging is a popular technique for the diagnosis of lung diseases \cite{sluimer2006computer,kamble2020review}. In practice, segmenting different organs and lesions from chest CT slices can provide crucial information for doctors to diagnose and quantify lung diseases \cite{gordaliza2018unsupervised}. 
Recently, many works have been provided and obtained promising performances. These algorithms often employ a classifier with extracted features for nodule segmentation  in chest CT. For example, Keshani \etal~\cite{keshani2013lung} utilized the support vector machine (SVM) classifier to detect the lung nodule from CT slices. Shen \etal~\cite{shen2015automated} presented an automated lung segmentation system based on bidirectional chain code to improve the performance. 
However, the similar visual appearances of nodules and background makes it difficult for extracting the nodule regions. 
To overcome this issue, several deep learning algorithms have been proposed to learn a powerful visual representations~\cite{wang2017central,jin2018ct,jiang2018multiple}. For instance, Wang \etal~\cite{wang2017central} developed a central focused convolutional neural network to segment lung nodules from heterogeneous CT slices. Jin \etal~\cite{jin2018ct} utilized GAN-synthesized data to improve the training of a discriminative model for pathological lung segmentation. Jiang \etal~\cite{jiang2018multiple} designed two deep networks to segment lung tumors from CT slices by adding multiple residual streams of varying resolutions. Wu \etal~\cite{wu2020jcs} built an explainable COVID-19 diagnosis system by joint classification and segmentation.

\subsection{Annotation-Efficient Deep Learning}

In our work, we aim to segment the COVID-19 infection regions for quantifying and evaluating the disease progression. The (unsupervised) anomaly detection/segmentation could detect the anomaly region~\cite{Schlegl_2017,abnormal_survey,abnormal_isbi}, however, it can not identify whether the anomaly region is related to COVID-19. By contrast, based on the few labeled data, the semi-supervised model could identify the target region from other anomaly region, which is better suit for assessment of COVID-19.  Moreover, the transfer learning technique is another good  choice for dealing with limited data~\cite{transfer2016,Cheplygina2019}. But currently, the major issue for segmentation of COVID-19 infection is that there are already some public datasets (see~\cite{COVID19-AI-Collection}), but, being short of high quality pixel-level annotations. This problem will become more pronounced, even collecting large scale COVID-19 dataset, where the annotations are still expensive to acquire. Thus, our target is to utilize the limited annotation efficiently and leverage unlabeled data. Semi-supervised learning provides a more suitable solution to address this issue.

The main goal of semi-supervised learning (SSL) is to improve model performance using a limited number of labeled data and a large amount of unlabeled data \cite{zhou2019collaborative}. Currently, there is increasing focus on training deep neural network using the SSL strategy \cite{VanEngelen2020}. These methods often optimize a supervised loss on labeled data along with an unsupervised loss imposed on either unlabeled data \cite{lee2013pseudo} or both the labeled and unlabeled data \cite{laine2016temporal,rasmus2015semi}. Lee \etal~\cite{lee2013pseudo} provided to utilize a cross-entropy loss by computing on the pseudo labels of unlabeled data, which is considered as an additional supervision loss. In summary, existing deep SSL algorithms regularize the network by enforcing smooth and consistent classification boundaries, which are robust to a random perturbation \cite{rasmus2015semi}, and other approaches enrich the supervision signals by exploring the knowledge learned, \eg, based on the temporally ensembled prediction \cite{laine2016temporal} and pseudo label \cite{lee2013pseudo}. 
In addition, semi-supervised learning has been widely applied in medical segmentation task, where a frequent issue is the lack of pixel-level labeled data, even when large scale set of unlabeled image could be available~\cite{Cheplygina2019,Tajbakhsh2020}. For example, Nie \etal~\cite{nie2018asdnet} proposed an attention-based semi-supervised deep network for pelvic organ segmentation, in which a semi-supervised region-attention loss is developed to address the insufficient data issue for training deep learning models. Cui \etal~\cite{Cui_IPMI} modified a mean teacher framework for the task
of stroke lesion segmentation in MR images. Zhao \etal~\cite{zhoa_miccai_semi} proposed a semi-supervised segmentation method based on a self-ensemble architecture and a random patch-size training strategy. Different from these works, our semi-supervised framework is based on a random sampling strategy for progressively enlarging the training set with unlabeled data.

\begin{figure*}[t!]
	\centering
    \small
	\begin{overpic}[width=\textwidth]{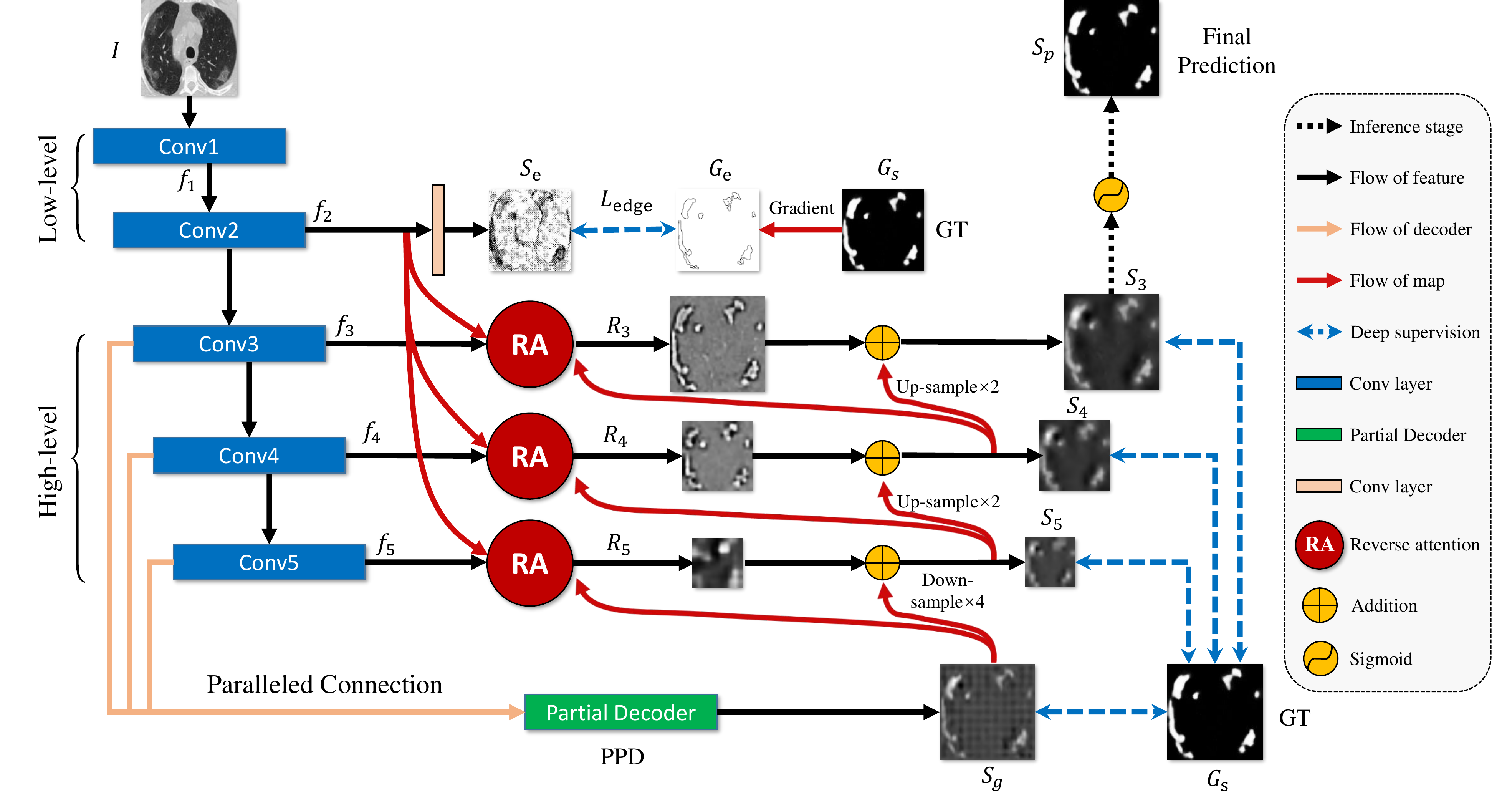}
    \end{overpic}
	\caption{\small The architecture of our proposed \ourmodel~model, which consists of three reverse attention (RA) modules connected to the paralleled partial decoder (PPD). See \secref{sec:ourmodel} for details.}
    \label{fig:Framework}
\end{figure*}

\subsection{Artificial Intelligence for COVID-19}

Artificial intelligence  technologies have been employed in a large number of applications against COVID-19 \cite{shi2020review,Dong2020review,9086482,9090149}. 
Joseph \etal~\cite{bullock2020mapping} categorized these applications into three scales, including patient scale (\eg, medical imaging for diagnosis \cite{wang2020deep,chen2020deep}), molecular scale (\eg, protein structure prediction \cite{senior2018alphafold}), and societal scale (\eg, epidemiology \cite{hu2020artificial}).
In this work, we focus on patient scale applications \cite{wang2020deep,chen2020deep,xu2020deep,gozes2020rapid,tang2020severity,shan+2020lung,shi2020large}, especially those based on CT slices. % Cite more work?
For instance, Wang \etal~\cite{wang2020deep} proposed a modified inception neural network \cite{szegedy2015going} for classifying COVID-19 patients and normal controls. Instead of directly training on complete CT images, they trained the network on the regions of interest, which are identified by two radiologists based on the features of pneumonia.
Chen \etal~\cite{chen2020deep} collected 46,096 CT image slices from COVID-19 patients and control patients of other disease. The CT images collected were utilized to train a U-Net++ \cite{zhou2018unetplus} for identifying COVID-19 patients. Their experimental results suggest that the trained model performs comparably with expert radiologists in terms of COVID-19 diagnosis.
In addition, other network architectures have also been considered in developing AI-assisted COVID-19 diagnosis systems. Typical examples include ResNet, used in \cite{xu2020deep}, and U-Net \cite{ronneberger2015u}, used in \cite{gozes2020rapid}.
Finally, deep learning has been employed to segment the infection regions in lung CT slices so that the resulting quantitative features can be utilized for severity assessment \cite{tang2020severity}, large-scale screening \cite{shi2020large}, and lung infection quantification \cite{Chaganti2020,shan+2020lung,bullock2020mapping} of COVID-19.

%Moreover, several methods have been proposed to automatically detect and quantifies abnormal tomographic patterns commonly present in COVID-19 \cite{Chaganti2020}, 

\begin{figure}[thp!]
	\centering
	\small
	\begin{overpic}[width=.9\columnwidth]{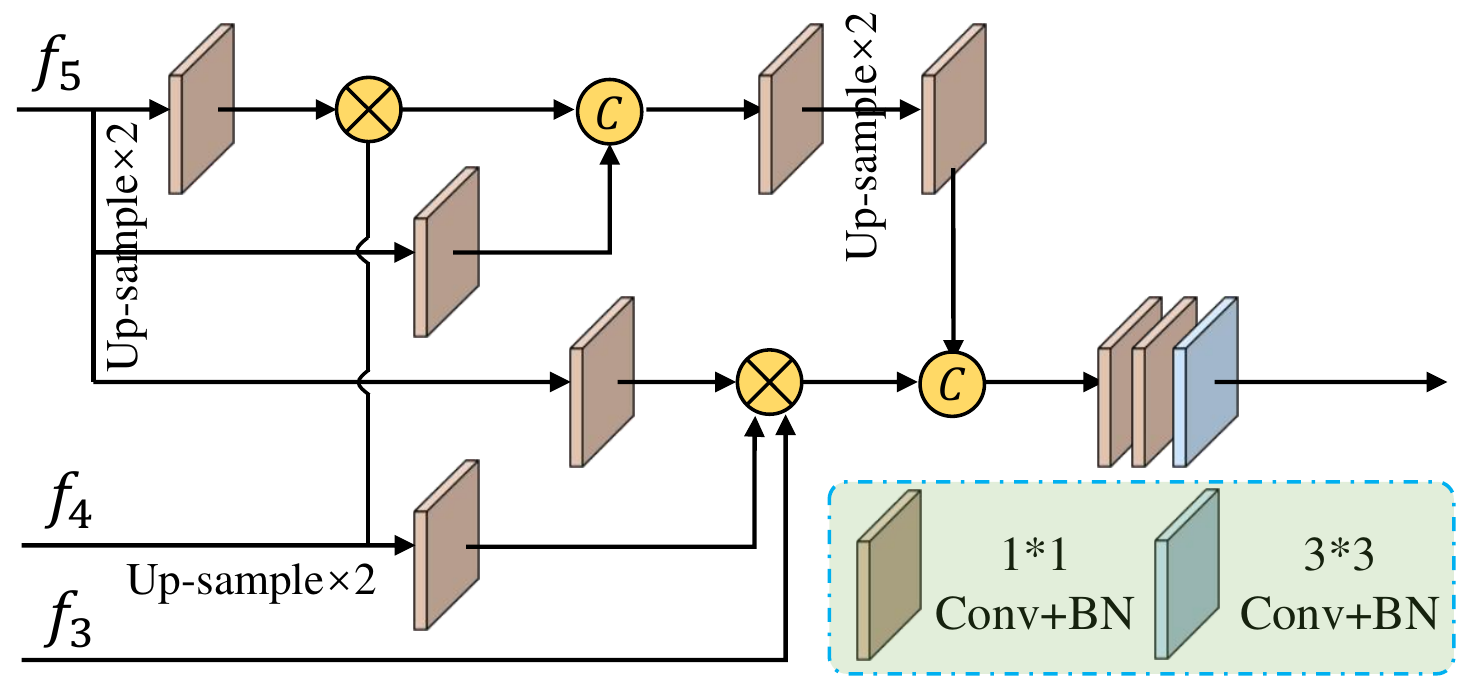}
	\end{overpic}
	\caption{\small Paralleled partial decoder is utilized to generate the global map.}
	\label{fig:PPD}
\end{figure}

\begin{figure}[t]
	\centering
    \small
	\begin{overpic}[width=\columnwidth]{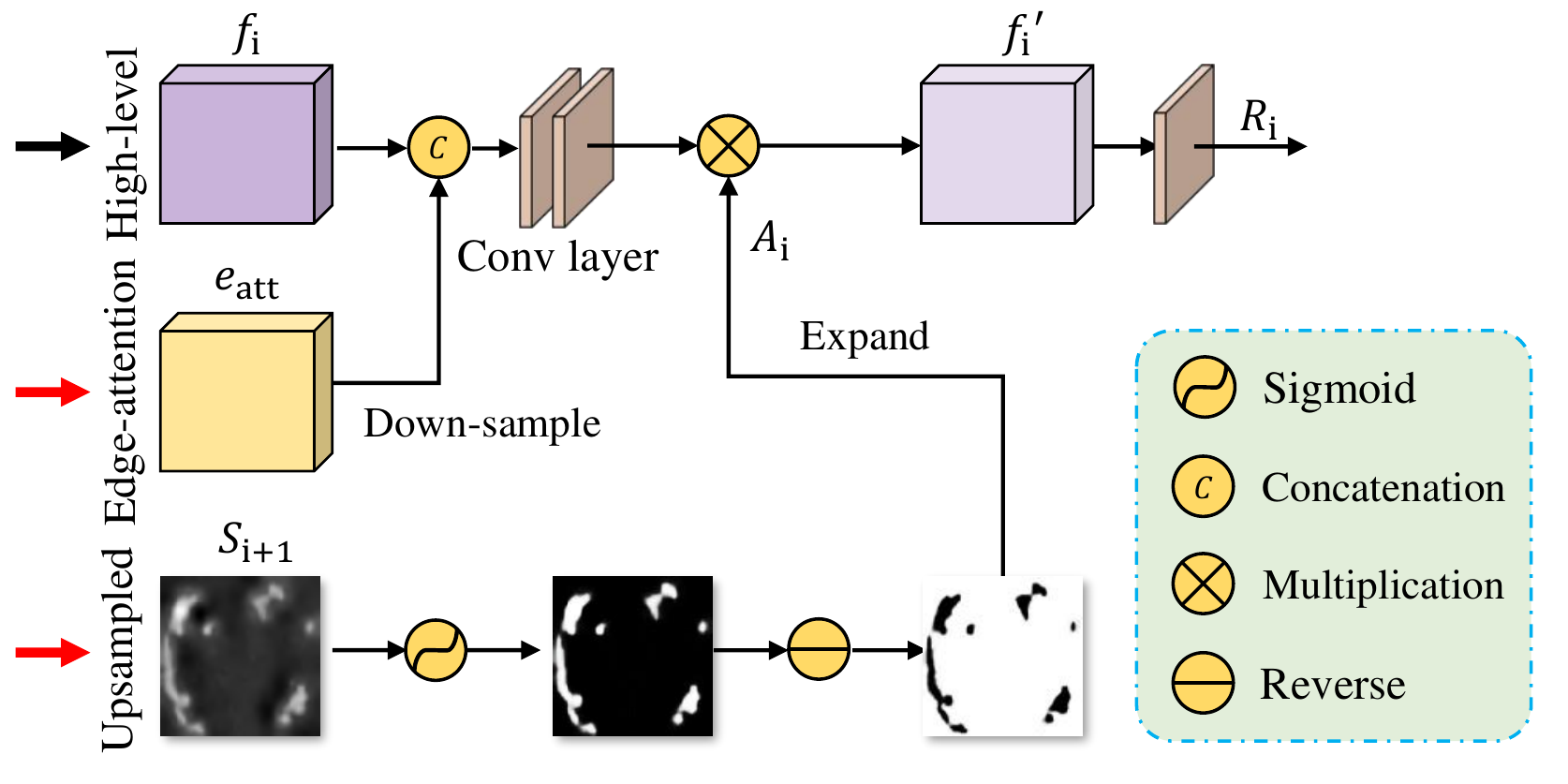}
    \end{overpic}
	\caption{\small Reverse attention module is utilized to implicitly learning edge features.}
    \label{fig:RA}
\end{figure}

\section{Proposed Method}
 
In this section, we first provide details of our \ourmodel~in terms of network architecture, core network components, and loss function. We then present the semi-supervised version of \ourmodel~and clarify how to use a semi-supervised learning framework to enlarge the limited number of training samples for improving the segmentation accuracy. We also show an extension of our framework for the multi-class labeling of different types of lung infections. Finally, we provide the implementation details.

\subsection{Lung Infection Segmentation Network (\ourmodel)}\label{sec:ourmodel}

%\noindent\textbf{Network Architecture:}
\noindent\textbf{Overview of Network:}
The architecture of our \ourmodel~is shown in Fig.~\ref{fig:Framework}. As can be observed, CT images are first fed to two convolutional layers to extract high-resolution, semantically weak (\ie, low-level) features. Herein, we add an edge attention module to \textit{explicitly} improve the representation of objective region boundaries.
% \comment{Please check the places marked by 'xxx'.}
Then, the low-level features $f_2$ obtained are fed to three convolutional layers for extracting the high-level features, which are used for two purposes. First, we utilize a {\textit{parallel partial decoder}} (PPD) to aggregate these features and generate a global map $S_g$ for the coarse localization of lung infections. Second, these features combined with $f_2$ are fed to multiple {\textit{reverse attention}} (RA) modules under the guidance of the $S_g$.
It is worth noting that the RA modules are organized in a cascaded fashion. For instance, as shown in Fig.~\ref{fig:Framework}, $R_4$ relies on the output of another RA $R_5$.
Finally, the output of the last RA, \ie, $S_3$, is fed to a \emph{Sigmoid} activation function for the final prediction of lung infection regions.
We now detail the key components of \ourmodel~ and our loss function.

%\noindent
\textbf{Edge Attention Module}: 
Several works have shown that edge information can provide useful constraints to guide feature extraction for segmentation \cite{zhao2019egnet,wu2019stacked,ETNet}. Thus, considering that the low-level features (\eg, $f_2$ in our model) preserve some sufficient edge information, we feed the low-level feature $f_2$ with moderate resolution to the proposed \textbf{\textit{edge attention}} (EA) module to explicitly learn an edge-attention representation. Specifically, the feature $f_2$ is fed to a convolutional layer with one filter to produce the edge map. Then, we can measure the dissimilarity of the EA module between the produced edge map and the edge map $G_e$ derived from the ground-truth (GT), which is constrained by the standard Binary Cross Entropy (BCE) loss function:
%
% \comment{1. Need to add loss for edge module. 2. Should we move this loss section to the totally Loss Function. 3. How to generate edge map?}
%
\begin{equation}\label{equ:edge_sup}
\mathcal{L}_{edge} = - \sum_{x=1}^{w} \sum_{y=1}^{h} [G_{e}log(S_{e})+(1-G_{e})log(1-S_{e})],
\end{equation}
where $(x,y)$ are the coordinates of each pixel in the predicted edge map $S_e$ and edge ground-truth map $G_e$. The $G_e$ is calculated using the gradient of the ground-truth map $G_s$. Additionally, $w$ and $h$ denote the width and height of corresponding map, respectively.

\textbf{Parallel Partial Decoder:}
Several existing medical image segmentation networks segment interested organs/lesions using all high- and low-level features in the encoder branch~\cite{ronneberger2015u,zhou2018unetplus,MNet,Gu2019,Zhang2019_AGNet,Fabian2020Automated}.
However, Wu~\etal\cite{wu2019cascaded} pointed out that, compared with high-level features, low-level features demand more computational resources due to larger spatial resolutions, but contribute less to the performance.
% %
% \comment{The motivation should be double checked. Not clear why PD can resolve problems raised by Wu \etal. In addition, the high-level features of U-Net share the same spatial resolution with the low-level features.}
% %
Inspired by this observation, we propose to only aggregate high-level features with a \textbf{\textit{parallel partial decoder}} component, illustrated in Fig.~\ref{fig:PPD}.
Specifically, for an input CT image $I$, we first extract two sets of low-level features $\{f_i, i = 1,2 \}$ and three sets of high-level features $\{f_i, i = 3,4,5 \}$ using the first five convolutional blocks of Res2Net~\cite{gao2019res2net}.
% Res2Net-based~\cite{pami20Res2net} backbone network.
% We then divide $f_i$ features into low-level features $\{f_i,i = 1,2\}$ and high-level features $\{f_i, i = 3,4,5\}$.
We then utilize the partial decoder $p_d (\cdot)$~\cite{wu2019cascaded}, a novel decoder component, to aggregate the high-level features with a paralleled connection.
The partial decoder yields a coarse global map $S_g = p_d(f_3,f_4, f_5)$, which then serves as global guidance in our RA modules.

\textbf{Reverse Attention Module:}
In clinical practice, clinicians usually segment lung infection regions via a two-step procedure, by roughly localizing the infection regions and then accurately labeling these regions by inspecting the local tissue structures.
Inspired by this procedure, we design \ourmodel~using two different network components that act as a rough locator and a fine labeler, respectively.
First, the PPD acts as the rough locator and yields a global map $S_g$, which provides the rough location of lung infection regions, without structural details (see Fig.~\ref{fig:Framework}).
Second, we propose a progressive framework, acting as the fine labeler, to mine discriminative infection regions in an erasing manner~\cite{wei2017object,chen2018reverse}.
Specifically, instead of simply aggregating features from all levels \cite{chen2018reverse}, we propose to adaptively learn the \textit{\textbf{reverse attention}} in three parallel high-level features.
Our architecture can sequentially exploit complementary regions and details by erasing the estimated infection regions from high-level side-output features, where the existing estimation is up-sampled from the deeper layer.

We obtain the output RA features $R_i$ by multiplying (element-wise $\odot$) the fusion of high-level side-output features $\{f_i, i=3,4,5\}$ and edge attention features $e_{att}=f_2$ with RA weights $A_i$, \ie,
%\vspace{-0.4cm}
\begin{equation} \label{equ:RA}
R_i = \mathcal{C} (f_i,~Dow({e}_{att})) \odot A_i,
\end{equation}
where Dow($\cdot$) denotes the down-sampling operation, $\mathcal{C}(\cdot)$ denotes the concatenation operation follow by two 2-D convolutional layers with 64 filters.

\begin{figure*}[t!]
	\centering
    \small
	\begin{overpic}[width=.88\textwidth]{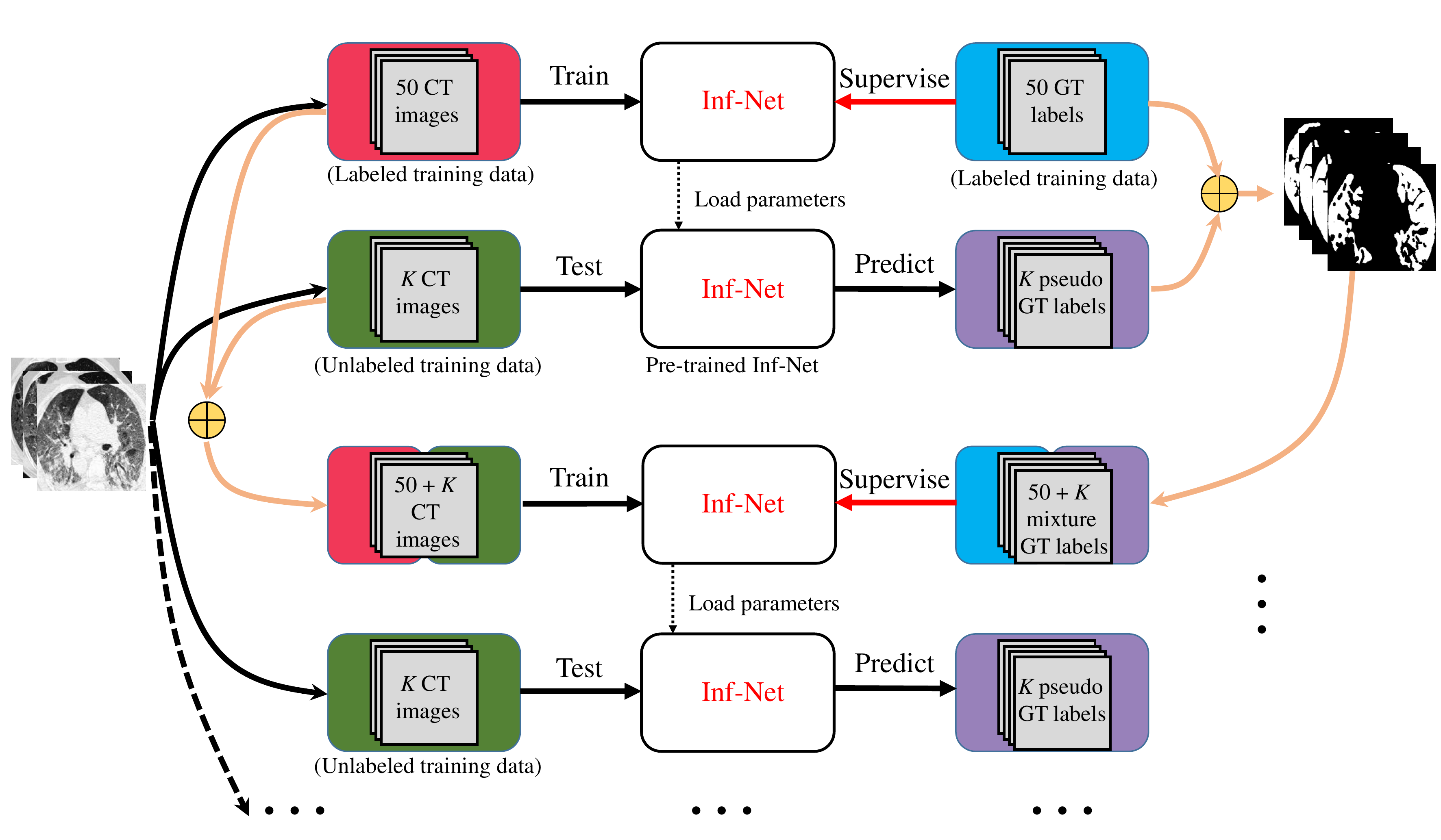}
    \end{overpic}
	\caption{\small Overview of the proposed \textit{Semi-supervised \ourmodel} framework. Please refer to \secref{sec:semi-inf-net} for more details.} 
    \label{fig:Pipline}
\end{figure*}

The RA weight $A_i$ is de-facto for salient object detection in the computer vision community~\cite{chen2018reverse}, and it is defined as:
\begin{equation} \label{WeightAttention}
A_i = \mathcal{E}(\circleddash(\sigma(\mathcal{P}(S_{i+1})))),
\end{equation}
where $\mathcal{P}(\cdot)$ denotes an up-sampling operation, $\sigma(\cdot)$ is a \emph{Sigmoid} activation function, and $\circleddash(\cdot)$ is a reverse operation subtracting the input from matrix ${E}$, in which all the elements are $1$.
Symbol $\mathcal{E}$ denotes expanding a single channel feature to 64 repeated tensors, which involves reversing each channel of the candidate tensor in Eq.~\eqref{equ:RA}.
Details of this procedure are shown in Fig.~\ref{fig:RA}.
% \comment{Currently, there is no illustrations for PD and RA.}
It is worth noting that the erasing strategy driven by RA can eventually refine the imprecise and coarse estimation into an accurate and complete prediction map.

% \noindent
\textbf{Loss Function:}
As mentioned above in Eq.~\eqref{equ:edge_sup}, we propose the loss function $\mathcal{L}_{edge}$ for edge supervision. Here, we define our loss function $\mathcal{L}_{seg}$ as a combination of a weighted IoU loss $\mathcal{L}_{IoU}^w$ and a weighted binary cross entropy (BCE) loss $\mathcal{L}_{BCE}^w$ for each segmentation supervision, \ie,
\begin{equation}\label{eq:loss_raw}
\mathcal{L}_{seg} = \mathcal{L}_{IoU}^w + \lambda \mathcal{L}_{BCE}^w ,
\end{equation}
where $\lambda$ is the weight, and set to 1 in our experiment.
The two parts of $\mathcal{L}_{seg}$ provide effective global (image-level) and local (pixel-level) supervision for accurate segmentation.
Unlike the standard IoU loss, which has been widely adopted in segmentation tasks, the weighted IoU loss increases the weights of hard pixels to highlight their importance.
In addition, compared with the standard BCE loss, $\mathcal{L}_{BCE}^w$ puts more emphasis on hard pixels rather than assigning all pixels equal weights.
The definitions of these losses are the same as in \cite{qin2019basnet,wei2019f3net} and their effectiveness has been validated in the field of salient object detection.
Note that the Correntropy-induced loss functions \cite{chen2016efficient,liangjun2018correntropy} can be employed here for improving the robustness.

Finally, we adopt deep supervision for the three side-outputs (\ie, $S_3$, $S_4$, and $S_5$) and the global map $S_g$.
Each map is up-sampled (\eg, $S_3^{up}$) to the same size as the object-level segmentation ground-truth map $G_s$.
Thus, the total loss in Eq.~\eqref{eq:loss_raw} is extended to
\begin{equation}
\mathcal{L}_{total} = \mathcal{L}_{seg} (G_s, S_g^{up}) + \mathcal{L}_{edge} + \sum_{i=3}^{i=5} \mathcal{L}_{seg} (G_s, S_i^{up}).
\end{equation}

\subsection{Semi-Supervised \ourmodel}\label{sec:semi-inf-net}
 
Currently, there is very limited number of CT images with segmentation annotations, since manually segmenting lung infection regions are difficult and time-consuming, and the disease is at an early stage of outbreak. %, indicating considerable difficulties in collecting a large-scale dataset.
To resolve this issue, we improve \ourmodel~using a semi-supervised learning strategy, which leverages a large number of unlabeled CT images to effectively augment the training dataset.
An overview of our semi-supervised learning framework is shown in Fig.~\ref{fig:Pipline}.
Our framework is mainly inspired by the work in \cite{mittal2019parting}, which is based on a random sampling strategy for progressively enlarging the training dataset with unlabeled data.
% to add a certain set of data from the unlabeled data to the labeled set with predicted pseudo labels. 
%\comment{Please check the places marked by 'xxx'.}
Specifically, we generate the pseudo labels for unlabeled CT images using the procedure described in Algorithm~\ref{alg:semi}. The resulting CT images with pseudo labels are then utilized to train our model using a two-step strategy detailed in Section~\ref{sec:imp-details}.
 
\renewcommand{\algorithmicrequire}{\textbf{Input:}}
\renewcommand{\algorithmicensure}{\textbf{Output:}}
\begin{algorithm}[t]
	\caption{Semi-Supervised \ourmodel} 
	\label{alg:semi}
	\begin{algorithmic}[1]
		\REQUIRE{Labeled training data $\mathcal{D}_{\text{Labeled}}$ and unlabeled training data $\mathcal{D}_{\text{Unlabeled}}$}
		\ENSURE{Trained \ourmodel~$\mathcal{M}$}
		\STATE{Construct a training dataset $\mathcal{D}_{\text{Training}}$ using all the labeled CT images from $\mathcal{D}_{\text{Labeled}}$}
		\STATE{Train our model $\mathcal{M}$ using $\mathcal{D}_{\text{Training}}$}
		\REPEAT
		\STATE{Perform testing using the trained model $\mathcal{M}$ and $K$ CT images randomly selected from $\mathcal{D}_{\text{Unlabeled}}$, which yields network-labeled data $\mathcal{D}_{\text{Net-labeled}}$, consisting of $K$ CT images with pseudo labels}
		\STATE{Enlarge the training dataset using $\mathcal{D}_{\text{Net-labeled}}$, \ie, $\mathcal{D}_{\text{Training}} = \mathcal{D}_{\text{Training}} \cup \mathcal{D}_{\text{Net-labeled}}$}
		\STATE{Remove the $K$ testing CT images from $\mathcal{D}_{\text{Unlabeled}}$}
		\STATE{Fine-tune $\mathcal{M}$ using $\mathcal{D_{\text{Training}}}$}
		\UNTIL{$\mathcal{D}_{\text{Unlabeled}}$ is empty}
		\RETURN{Trained model $\mathcal{M}$}
	\end{algorithmic}
\end{algorithm} 

The advantages of our framework, called \textit{Semi-}\ourmodel,~lie in two aspects.
First, the training and selection strategy is simple and easy to implement. It does not require measures to assess the predicted label, and it is also threshold-free.
% \footnote{The threshold should be assigned to select top x\% data in previous semi-supervised learning methods.}.
Second, this strategy can provide more robust performance than other semi-supervised learning methods and prevent over-fitting. This conclusion is confirmed by recently released studies~\cite{mittal2019parting}.

%\comment{Please check the places marked by 'xxx'.}

\begin{figure}[!t]
	\centering
    \small
	\begin{overpic}[width=.9\columnwidth]{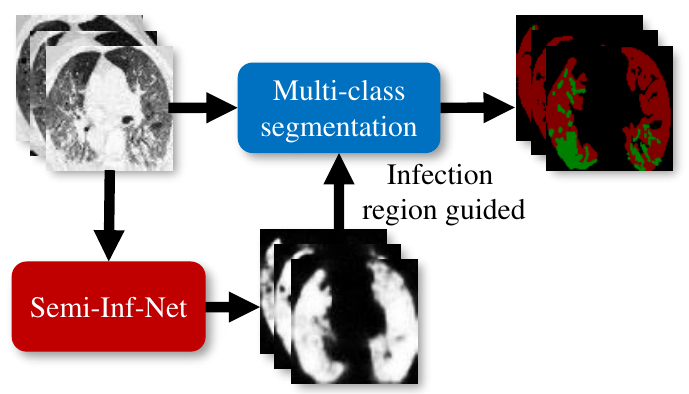}
    \end{overpic}
	\caption{Illustration of infection region guided multi-class segmentation for multi-class labeling task. We feed both the infection segmentation results provided by \ourmodel~and the CT images into FCN8s (or Multi-class U-Net) for improving the accuracy of multi-class infection labeling.}
    \label{fig:ComparedModel}
\end{figure}

% \subsection{Extension to Multi-Label Segmentation}
\subsection{Extension to Multi-Class Infection Labeling}

Our \textit{Semi-}\ourmodel~is a powerful tool that can provide crucial information for evaluating overall lung infections. However, we are aware that, in a clinical setting, in addition to the overall evaluation, clinicians might also be interested in the quantitative evaluation of different kinds of lung infections, \eg, GGO and consolidation.
% For instance, COVID-19 lung infections can be categorized into two major types, including round-glass and consolidation.
Therefore, we extend \textit{Semi-}\ourmodel~to a multi-class lung infection labeling framework so that it can provide richer information for the further diagnosis and treatment of COVID-19.
The extension of \textit{Semi-}\ourmodel~is based on an infection region guided multi-class labeling framework, which is illustrated in Fig.~\ref{fig:ComparedModel}.
Specifically, we utilize the infection segmentation results provided by \textit{Semi-}\ourmodel~to guide the multi-class labeling of different types of lung infections.
For this purpose, we feed both the infection segmentation results and the corresponding CT images to a multi-class segmentation network, \eg, FCN8s~\cite{long2015fully}, or U-Net~\cite{ronneberger2015u}.
This framework can take full advantage of the infection segmentation results provided by \textit{Semi-}\ourmodel~and effectively improve the performance of multi-class infection labeling. 

\subsection{Implementation Details}\label{sec:imp-details}
Our model is implemented in PyTorch, and is accelerated by an NVIDIA TITAN RTX GPU. We describe the implementation details as follows.

\textbf{Pseudo label generation:}
We generate pseudo labels for unlabeled CT images using the protocol described in Algorithm~\ref{alg:semi}. 
The number of randomly selected CT images is set to 5, \ie, $K=5$.
For 1600 unlabeled images, we need to perform 320 iterations with a batch size of 16. The entire procedure takes about 50 hours to complete.

\textbf{Semi-supervised \ourmodel:}
Before training, we uniformly resize all the inputs to $352 \times 352$.
We train \ourmodel~using a multi-scale strategy \cite{wu2019stacked}. Specifically, we first re-sample the training images using different scaling ratios, \ie, $\{0.75, 1, 1.25\}$, and then train \ourmodel~using the re-sampled images, which improves the generalization of our model.
%We train \ourmodel~using a multi-scale strategy with different scale ratios $\{0.75, 1, 1.25\}$.
% \comment{What is so-called multi-scale training strategy?}
The Adam optimizer is employed for training and the learning rate is set to $1e-4$.
Our training phase consists of two steps: (i) Pre-training on 1600 CT images with pseudo labels, which takes $\sim$180 minutes to converge over 100 epochs with a batch size of 24. (ii) Fine-tuning on 50 CT images with the ground-truth labels, which takes $\sim$15 minutes to converge over 100 epochs with a batch size of 16.
For a fair comparison, the training procedure of \textbf{\textit{Inf-Net}} follows the same setting described in the second step.
% maintains the same hyper-parameters with those described above, excepting for training with only 50 labeled images.

\textbf{\textit{Semi-}\ourmodel+Multi-class segmentation:}
%\comment{Check the name of model}
For Multi-class segmentation network, we are not constrained to specific choice of the segmentation network, and herein FCN8s~\cite{long2015fully} and U-Net~\cite{ronneberger2015u} are used as two backbones.  We resize all the inputs to $512 \times 512$ before training. %and data augmentation.
The network is initialized by a uniform Xavier, and trained using an SGD optimizer with a learning rate of $1e-10$, weight decay of $5e-4$, and momentum of $0.99$.
The entire training procedure takes about 45 minutes to complete.

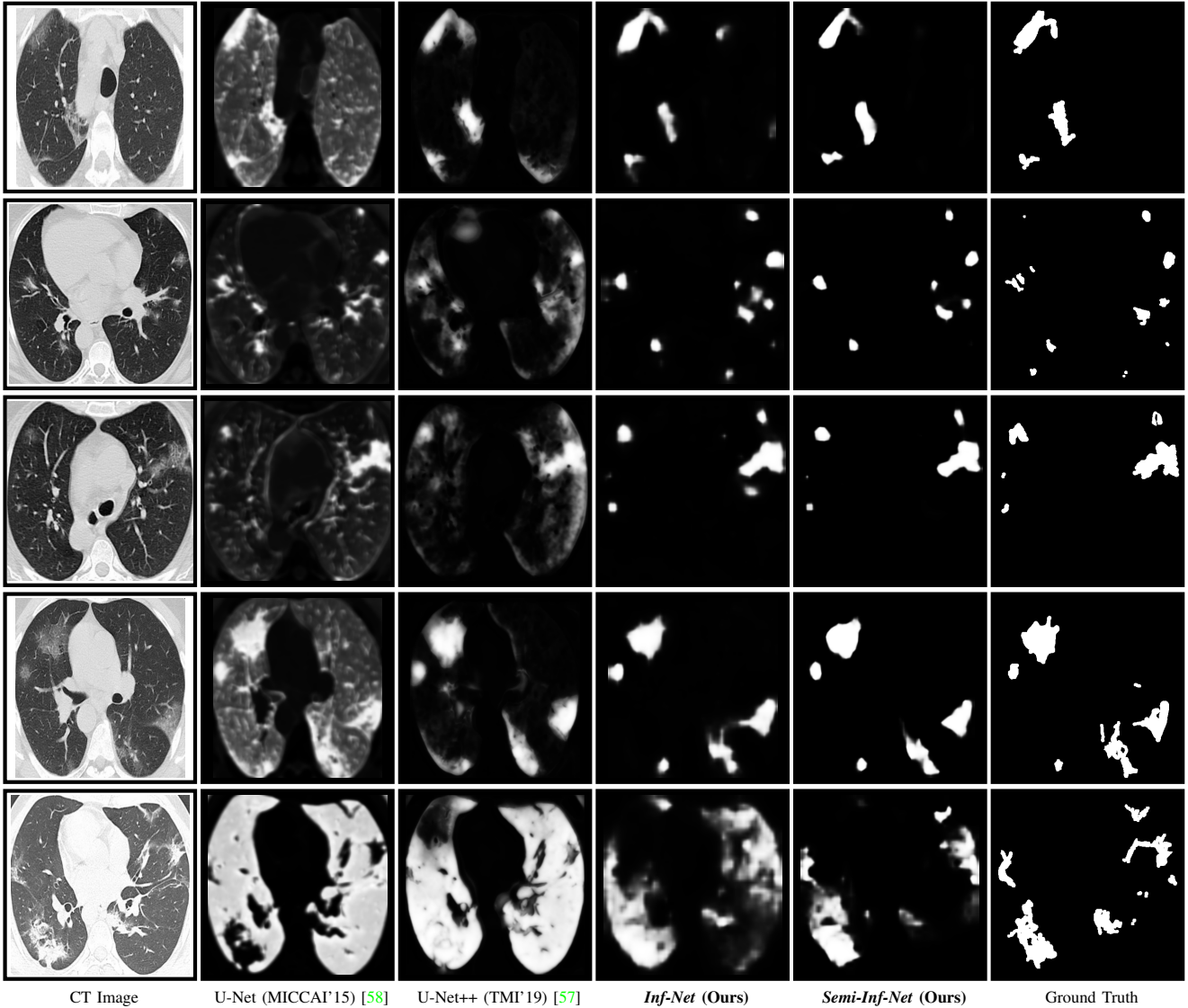
\begin{figure*}[t]
	\centering
	
	\setlength{\iwidth}{0.155\textwidth}
	\setlength{\iheight}{0.99\iwidth}
	\setlength{\tabcolsep}{-1pt}
	\setlength{\rowspacing}{-1.5pt}

	\newcommand{\scaleRatio}{0.95}
	\newcommand{\picLocationX}{.5}
	\newcommand{\picLocationY}{.5}
	\def \filePath {Imgs/visual_results}
	
	% Command to include pictures
	\newcommand\myPic[8]{
		\begin{tikzpicture} % [baseline,trim left]
		% foreground
%		\setlength{\iwidth}{#8}
		%\fill (0.8\iwidth, 0.0\iheight)[black, ultra thick] rectangle(\iwidth, 0.05\iheight); % Remove the numbers in background
		\draw (0.0\iwidth, 0.0\iheight)[#5, ultra thick] rectangle(\iwidth, \iheight);
		%		\node[fancylabel, right=-6pt, text=black, yshift=3pt] at (0.0\iwidth, 1.0\iheight) {\textbf{\scriptsize{#6}}}; % Options for labels: Please refer to packages.tex
%		\node[fancylabel, right=-5pt, text=black, yshift=2pt] at (0.0\iwidth, 1.0\iheight) {\textbf{\scriptsize{#6}}}; % Options for labels: Please refer to packages.tex
		
		\begin{pgfonlayer}{background}
		\fill (0.0\iwidth, 0.0\iheight)[black, ultra thick] rectangle(\iwidth, \iheight);
		\clip (0.0\iwidth, 0.0\iheight) rectangle (\iwidth, \iheight);
%		\node at (#3\iwidth, #4\iheight) { \includegraphics[height=#2\iheight, angle=#7]{\filePath/#1} };
%		\node at (#3\iwidth, #4\iheight) { \includegraphics[height=#2\iheight, angle=#7]{\filePath/#1} };
		\node at (#3\iwidth, #4\iheight) { \includegraphics[width=#2\iheight, angle=270]{\filePath/#1} };
		\end{pgfonlayer}
		\end{tikzpicture}
	}

	% Command to include pictures
	\newcommand\myPicNew[8]{
		\begin{tikzpicture} % [baseline,trim left]
		% foreground
		%		\setlength{\iwidth}{#8}
		%\fill (0.8\iwidth, 0.0\iheight)[black, ultra thick] rectangle(\iwidth, 0.05\iheight); % Remove the numbers in background
		\draw (0.0\iwidth, 0.0\iheight)[#5, ultra thick] rectangle(\iwidth, \iheight);
		%		\node[fancylabel, right=-6pt, text=black, yshift=3pt] at (0.0\iwidth, 1.0\iheight) {\textbf{\scriptsize{#6}}}; % Options for labels: Please refer to packages.tex
		%		\node[fancylabel, right=-5pt, text=black, yshift=2pt] at (0.0\iwidth, 1.0\iheight) {\textbf{\scriptsize{#6}}}; % Options for labels: Please refer to packages.tex
		
		\begin{pgfonlayer}{background}
		\fill (0.0\iwidth, 0.0\iheight)[white, ultra thick] rectangle(\iwidth, \iheight);
		\clip (0.0\iwidth, 0.0\iheight) rectangle (\iwidth, \iheight);
		%		\node at (#3\iwidth, #4\iheight) { \includegraphics[height=#2\iheight, angle=#7]{\filePath/#1} };
		%		\node at (#3\iwidth, #4\iheight) { \includegraphics[height=#2\iheight, angle=#7]{\filePath/#1} };
		\node at (#3\iwidth, #4\iheight) { \includegraphics[width=#2\iheight, angle=270]{\filePath/#1} };
		\end{pgfonlayer}
		\end{tikzpicture}
	}
	
	% Usefule tools (column text and color bars)
		\newcommand\columnText[1]{
%	\begin{tikzpicture}[baseline,trim left]
%	\node[rotate=90] at (0.05\iwidth, 0.5\iheight) {{#1}};
%	\end{tikzpicture}
	}
	
	% For annotations
	\newcommand\voidAnn{}

%	\hrule % Reference line
	%	\centering
	
	\begin{tabular}{cccccc}
	%	% The usage of \myPic:  1. Picture name; 2. Scale ratio; 3. x shift; 4. y shift; 5. Color; 6. Annotations; 7. Label;
%	\myPicNew{GT/COVID-19_images/89}{\scaleRatio}{\picLocationX}{\picLocationY}{black}{}{90}{\iwidth} & \myPic{PredMap/Object-level_12/UNet_12-50epoch/COVID-19_12/89}{\scaleRatio}{\picLocationX}{\picLocationY}{black}{}{90}{\iwidth} & \myPic{PredMap/Object-level_12/UNetPlusPlus_12-50epoch/COVID-19_12/89}{\scaleRatio}{\picLocationX}{\picLocationY}{black}{}{90}{\iwidth} & \myPic{PredMap/Object-level_12/PraNet++_12-100epoch/COVID-19_12/89}{\scaleRatio}{\picLocationX}{\picLocationY}{black}{}{90}{\iwidth}  & \myPic{PredMap/Object-level_12/PraNet++_Semi_12-100epoch(50finetune-new)/COVID-19_12/89}{\scaleRatio}{\picLocationX}{\picLocationY}{black}{}{90}{\iwidth}  & \myPic{GT/COVID-19_12/89}{\scaleRatio}{\picLocationX}{\picLocationY}{black}{}{90}{\iwidth}
	
%	\\[\rowspacing] 
	\myPicNew{GT/COVID-19_images/97}{\scaleRatio}{\picLocationX}{\picLocationY}{black}{}{90}{\iwidth} & \myPic{PredMap/Object-level_12/UNet_12-50epoch/COVID-19_12/97}{\scaleRatio}{\picLocationX}{\picLocationY}{black}{}{90}{\iwidth} & \myPic{PredMap/Object-level_12/UNetPlusPlus_12-50epoch/COVID-19_12/97}{\scaleRatio}{\picLocationX}{\picLocationY}{black}{}{90}{\iwidth} & \myPic{PredMap/Object-level_12/PraNet++_12-100epoch/COVID-19_12/97}{\scaleRatio}{\picLocationX}{\picLocationY}{black}{}{90}{\iwidth}  & \myPic{PredMap/Object-level_12/PraNet++_Semi_12-100epoch(50finetune-new)/COVID-19_12/97}{\scaleRatio}{\picLocationX}{\picLocationY}{black}{}{90}{\iwidth}  & \myPic{GT/COVID-19_12/97}{\scaleRatio}{\picLocationX}{\picLocationY}{black}{}{90}{\iwidth}

	\\[\rowspacing] \myPicNew{GT/COVID-19_images/61}{\scaleRatio}{\picLocationX}{\picLocationY}{black}{}{90}{\iwidth} & \myPic{PredMap/Object-level_12/UNet_12-50epoch/COVID-19_12/61}{\scaleRatio}{\picLocationX}{\picLocationY}{black}{}{90}{\iwidth} & \myPic{PredMap/Object-level_12/UNetPlusPlus_12-50epoch/COVID-19_12/61}{\scaleRatio}{\picLocationX}{\picLocationY}{black}{}{90}{\iwidth} & \myPic{PredMap/Object-level_12/PraNet++_12-100epoch/COVID-19_12/61}{\scaleRatio}{\picLocationX}{\picLocationY}{black}{}{90}{\iwidth}  & \myPic{PredMap/Object-level_12/PraNet++_Semi_12-100epoch(50finetune-new)/COVID-19_12/61}{\scaleRatio}{\picLocationX}{\picLocationY}{black}{}{90}{\iwidth}  & \myPic{GT/COVID-19_12/61}{\scaleRatio}{\picLocationX}{\picLocationY}{black}{}{90}{\iwidth}
		
	\\[\rowspacing]
	\myPicNew{GT/COVID-19_images/91}{\scaleRatio}{\picLocationX}{\picLocationY}{black}{}{90}{\iwidth} & \myPic{PredMap/Object-level_12/UNet_12-50epoch/COVID-19_12/91}{\scaleRatio}{\picLocationX}{\picLocationY}{black}{}{90}{\iwidth} & \myPic{PredMap/Object-level_12/UNetPlusPlus_12-50epoch/COVID-19_12/91}{\scaleRatio}{\picLocationX}{\picLocationY}{black}{}{90}{\iwidth} & \myPic{PredMap/Object-level_12/PraNet++_12-100epoch/COVID-19_12/91}{\scaleRatio}{\picLocationX}{\picLocationY}{black}{}{90}{\iwidth}  & \myPic{PredMap/Object-level_12/PraNet++_Semi_12-100epoch(50finetune-new)/COVID-19_12/91}{\scaleRatio}{\picLocationX}{\picLocationY}{black}{}{90}{\iwidth}  & \myPic{GT/COVID-19_12/91}{\scaleRatio}{\picLocationX}{\picLocationY}{black}{}{90}{\iwidth}

	\\[\rowspacing]
	\myPicNew{GT/COVID-19_images/21}{\scaleRatio}{\picLocationX}{\picLocationY}{black}{}{90}{\iwidth} & \myPic{PredMap/Object-level_12/UNet_12-50epoch/COVID-19_12/21}{\scaleRatio}{\picLocationX}{\picLocationY}{black}{}{90}{\iwidth} & \myPic{PredMap/Object-level_12/UNetPlusPlus_12-50epoch/COVID-19_12/21}{\scaleRatio}{\picLocationX}{\picLocationY}{black}{}{90}{\iwidth} & \myPic{PredMap/Object-level_12/PraNet++_12-100epoch/COVID-19_12/21}{\scaleRatio}{\picLocationX}{\picLocationY}{black}{}{90}{\iwidth}  & \myPic{PredMap/Object-level_12/PraNet++_Semi_12-100epoch(50finetune-new)/COVID-19_12/21}{\scaleRatio}{\picLocationX}{\picLocationY}{black}{}{90}{\iwidth}  & \myPic{GT/COVID-19_12/21}{\scaleRatio}{\picLocationX}{\picLocationY}{black}{}{90}{\iwidth}

	\\[\rowspacing] \myPicNew{GT/COVID-19_images/38}{\scaleRatio}{\picLocationX}{\picLocationY}{black}{}{90}{\iwidth} & \myPic{PredMap/Object-level_12/UNet_12-50epoch/COVID-19_12/38}{\scaleRatio}{\picLocationX}{\picLocationY}{black}{}{90}{\iwidth} & \myPic{PredMap/Object-level_12/UNetPlusPlus_12-50epoch/COVID-19_12/38}{\scaleRatio}{\picLocationX}{\picLocationY}{black}{}{90}{\iwidth} & \myPic{PredMap/Object-level_12/PraNet++_12-100epoch/COVID-19_12/38}{\scaleRatio}{\picLocationX}{\picLocationY}{black}{}{90}{\iwidth}  & \myPic{PredMap/Object-level_12/PraNet++_Semi_12-100epoch(50finetune-new)/COVID-19_12/38}{\scaleRatio}{\picLocationX}{\picLocationY}{black}{}{90}{\iwidth}  & \myPic{GT/COVID-19_12/38}{\scaleRatio}{\picLocationX}{\picLocationY}{black}{}{90}{\iwidth}
		
%	\\[\rowspacing]
%	\myPicNew{GT/COVID-19_images/81}{\scaleRatio}{\picLocationX}{\picLocationY}{black}{}{90}{\iwidth} & \myPic{PredMap/Object-level_12/UNet_12-50epoch/COVID-19_12/81}{\scaleRatio}{\picLocationX}{\picLocationY}{black}{}{90}{\iwidth} & \myPic{PredMap/Object-level_12/UNetPlusPlus_12-50epoch/COVID-19_12/81}{\scaleRatio}{\picLocationX}{\picLocationY}{black}{}{90}{\iwidth} & \myPic{PredMap/Object-level_12/PraNet++_12-100epoch/COVID-19_12/81}{\scaleRatio}{\picLocationX}{\picLocationY}{black}{}{90}{\iwidth}  & \myPic{PredMap/Object-level_12/PraNet++_Semi_12-100epoch(50finetune-new)/COVID-19_12/81}{\scaleRatio}{\picLocationX}{\picLocationY}{black}{}{90}{\iwidth}  & \myPic{GT/COVID-19_12/81}{\scaleRatio}{\picLocationX}{\picLocationY}{black}{}{90}{\iwidth}
	
%	\\[\rowspacing]
%	\myPicNew{GT/COVID-19_images/83}{\scaleRatio}{\picLocationX}{\picLocationY}{black}{}{90}{\iwidth} & \myPic{PredMap/Object-level_12/UNet_12-50epoch/COVID-19_12/83}{\scaleRatio}{\picLocationX}{\picLocationY}{black}{}{90}{\iwidth} & \myPic{PredMap/Object-level_12/UNetPlusPlus_12-50epoch/COVID-19_12/83}{\scaleRatio}{\picLocationX}{\picLocationY}{black}{}{90}{\iwidth} & \myPic{PredMap/Object-level_12/PraNet++_12-100epoch/COVID-19_12/83}{\scaleRatio}{\picLocationX}{\picLocationY}{black}{}{90}{\iwidth}  & \myPic{PredMap/Object-level_12/PraNet++_Semi_12-100epoch(50finetune-new)/COVID-19_12/83}{\scaleRatio}{\picLocationX}{\picLocationY}{black}{}{90}{\iwidth}  & \myPic{GT/COVID-19_12/83}{\scaleRatio}{\picLocationX}{\picLocationY}{black}{}{90}{\iwidth}
	
	\\[2\rowspacing]
	\scriptsize{CT Image} & \scriptsize{U-Net (MICCAI'15)~\cite{ronneberger2015u}} & \scriptsize{U-Net++ (TMI'19)~\cite{zhou2018unetplus}} & \scriptsize{\textbf{\ourmodel~(Ours)}} & \scriptsize{\textbf{\textit{Semi-}\ourmodel~(Ours)}} & \scriptsize{Ground Truth}
	
\end{tabular}

	%	\fbox
%	\mybox{\includegraphics[trim={0.8in 7.75in 4.07in 0.74in},clip,width=0.45\textwidth,page=2,]{\figDir/tikz_figures.pdf}} % trim: left, down, right, up
%	\vspace{3pt}
	\caption{{Visual comparison of lung infection segmentation results.} \label{fig:visual_results_single}}
%	\vspace{-3pt}
\end{figure*}

 % \label{fig:visual_results_single}
\begin{figure*}[t]
	\centering
	
	\setlength{\iwidth}{0.155\textwidth}
	\setlength{\iheight}{0.99\iwidth}
	\setlength{\tabcolsep}{-1pt}
	\setlength{\rowspacing}{-1.5pt}

	\newcommand{\scaleRatio}{0.9}
	\newcommand{\picLocationX}{.5}
	\newcommand{\picLocationY}{.5}
	\def \filePath {Imgs/visual_results}
	
	% Command to include pictures
	\newcommand\myPic[9]{
		\begin{tikzpicture} % [baseline,trim left]
		% foreground
		\setlength{\iwidth}{#8}
		%\fill (0.8\iwidth, 0.0\iheight)[black, ultra thick] rectangle(\iwidth, 0.05\iheight); % Remove the numbers in background
		\draw (0.0\iwidth, 0.0\iheight)[#5, ultra thick] rectangle(\iwidth, \iheight);
		%		\node[fancylabel, right=-6pt, text=black, yshift=3pt] at (0.0\iwidth, 1.0\iheight) {\textbf{\scriptsize{#6}}}; % Options for labels: Please refer to packages.tex
%		\node[fancylabel, right=-5pt, text=black, yshift=2pt] at (0.0\iwidth, 1.0\iheight) {\textbf{\scriptsize{#6}}}; % Options for labels: Please refer to packages.tex
		
		\begin{pgfonlayer}{background}
		\fill (0.0\iwidth, 0.0\iheight)[#9, ultra thick] rectangle(\iwidth, \iheight);
		\clip (0.0\iwidth, 0.0\iheight) rectangle (\iwidth, \iheight);
%		\node at (#3\iwidth, #4\iheight) { \includegraphics[height=#2\iheight, angle=#7]{\filePath/#1} };
		\node at (#3\iwidth, #4\iheight) { \includegraphics[width=#2\iheight, angle=270]{\filePath/#1} };
		\end{pgfonlayer}
		\end{tikzpicture}
	}
	
	% Usefule tools (column text and color bars)
		\newcommand\columnText[1]{
%	\begin{tikzpicture}[baseline,trim left]
%	\node[rotate=90] at (0.05\iwidth, 0.5\iheight) {{#1}};
%	\end{tikzpicture}
	}
	
	% For annotations
	\newcommand\voidAnn{}

%	\hrule % Reference line
	%	\centering
\begin{tabular}{cccccc}

%	\\[\rowspacing]
	\myPic{GT/COVID-19_images/38}{\scaleRatio}{\picLocationX}{\picLocationY}{black}{}{90}{\iwidth}{white} & \myPic{PredMap/Instance-level/DeepLabv3Plus_Stride8_Baseline/COVID-19-Instance12/38}{\scaleRatio}{\picLocationX}{\picLocationY}{black}{}{90}{\iwidth}{black} & \myPic{PredMap/Instance-level/DeepLabv3Plus_Stride16_Baseline/COVID-19-Instance12/38}{\scaleRatio}{\picLocationX}{\picLocationY}{black}{}{90}{\iwidth}{black} & \myPic{PredMap/Instance-level/FCN8s_1100/COVID-19-Instance12/38}{\scaleRatio}{\picLocationX}{\picLocationY}{black}{}{90}{\iwidth}{black} & \myPic{PredMap/Instance-level/UNet_ParNet++_Semi/COVID-19-Instance12/38}{\scaleRatio}{\picLocationX}{\picLocationY}{black}{}{90}{\iwidth}{black} & \myPic{GT/COVID-19-Instance12/38}{\scaleRatio}{\picLocationX}{\picLocationY}{black}{}{90}{\iwidth}{black}
		
	\\[\rowspacing]
	\myPic{GT/COVID-19_images/84}{0.6}{\picLocationX}{\picLocationY}{black}{}{90}{\iwidth}{white} & \myPic{PredMap/Instance-level/DeepLabv3Plus_Stride8_Baseline/COVID-19-Instance12/84}{0.6}{\picLocationX}{\picLocationY}{black}{}{90}{\iwidth}{black} & \myPic{PredMap/Instance-level/DeepLabv3Plus_Stride16_Baseline/COVID-19-Instance12/84}{0.6}{\picLocationX}{\picLocationY}{black}{}{90}{\iwidth}{black} & \myPic{PredMap/Instance-level/FCN8s_1100/COVID-19-Instance12/84}{0.6}{\picLocationX}{\picLocationY}{black}{}{90}{\iwidth}{black} & \myPic{PredMap/Instance-level/UNet_ParNet++_Semi/COVID-19-Instance12/84}{0.6}{\picLocationX}{\picLocationY}{black}{}{90}{\iwidth}{black} & \myPic{GT/COVID-19-Instance12/84}{0.6}{\picLocationX}{\picLocationY}{black}{}{90}{\iwidth}{black}
	
	\\[\rowspacing] \myPic{GT/COVID-19_images/87}{0.95}{\picLocationX}{\picLocationY}{black}{}{90}{\iwidth}{white} & \myPic{PredMap/Instance-level/DeepLabv3Plus_Stride8_Baseline/COVID-19-Instance12/87}{0.95}{\picLocationX}{\picLocationY}{black}{}{90}{\iwidth}{black} & \myPic{PredMap/Instance-level/DeepLabv3Plus_Stride16_Baseline/COVID-19-Instance12/87}{0.95}{\picLocationX}{\picLocationY}{black}{}{90}{\iwidth}{black} & \myPic{PredMap/Instance-level/FCN8s_1100/COVID-19-Instance12/87}{0.95}{\picLocationX}{\picLocationY}{black}{}{90}{\iwidth}{black} & \myPic{PredMap/Instance-level/UNet_ParNet++_Semi/COVID-19-Instance12/87}{0.95}{\picLocationX}{\picLocationY}{black}{}{90}{\iwidth}{black} & \myPic{GT/COVID-19-Instance12/87}{0.95}{\picLocationX}{\picLocationY}{black}{}{90}{\iwidth}{black}

	\\[\rowspacing]
	\myPic{GT/COVID-19_images/88}{\scaleRatio}{\picLocationX}{\picLocationY}{black}{}{90}{\iwidth}{white} & \myPic{PredMap/Instance-level/DeepLabv3Plus_Stride8_Baseline/COVID-19-Instance12/88}{\scaleRatio}{\picLocationX}{\picLocationY}{black}{}{90}{\iwidth}{black} & \myPic{PredMap/Instance-level/DeepLabv3Plus_Stride16_Baseline/COVID-19-Instance12/88}{\scaleRatio}{\picLocationX}{\picLocationY}{black}{}{90}{\iwidth}{black} & \myPic{PredMap/Instance-level/FCN8s_1100/COVID-19-Instance12/88}{\scaleRatio}{\picLocationX}{\picLocationY}{black}{}{90}{\iwidth}{black} & \myPic{PredMap/Instance-level/UNet_ParNet++_Semi/COVID-19-Instance12/88}{\scaleRatio}{\picLocationX}{\picLocationY}{black}{}{90}{\iwidth}{black} & \myPic{GT/COVID-19-Instance12/88}{\scaleRatio}{\picLocationX}{\picLocationY}{black}{}{90}{\iwidth}{black}
	
	\\[\rowspacing]
	\myPic{GT/COVID-19_images/99}{0.84}{\picLocationX}{\picLocationY}{black}{}{90}{\iwidth}{white} & \myPic{PredMap/Instance-level/DeepLabv3Plus_Stride8_Baseline/COVID-19-Instance12/99}{0.84}{\picLocationX}{\picLocationY}{black}{}{90}{\iwidth}{black} & \myPic{PredMap/Instance-level/DeepLabv3Plus_Stride16_Baseline/COVID-19-Instance12/99}{0.84}{\picLocationX}{\picLocationY}{black}{}{90}{\iwidth}{black} & \myPic{PredMap/Instance-level/FCN8s_1100/COVID-19-Instance12/99}{0.84}{\picLocationX}{\picLocationY}{black}{}{90}{\iwidth}{black} & \myPic{PredMap/Instance-level/UNet_ParNet++_Semi/COVID-19-Instance12/99}{0.84}{\picLocationX}{\picLocationY}{black}{}{90}{\iwidth}{black} & \myPic{GT/COVID-19-Instance12/99}{0.84}{\picLocationX}{\picLocationY}{black}{}{90}{\iwidth}{black}
	
	\\[2\rowspacing]
	\scriptsize{CT Image} & \scriptsize{DeepLabV3+ (stride = 8)} & \scriptsize{DeepLabV3+ (stride = 16)} & \scriptsize{FCN8s} & \scriptsize{\textit{Semi-}\ourmodel~\& MC} & \scriptsize{Ground Truth}
\end{tabular}

	\caption{{Visual comparison of multi-class lung infection segmentation results, where the red and green labels indicate the GGO and consolidation, respectively.}
	\label{fig:visual_results_multi}}
%	\vspace{-3pt}
\end{figure*}
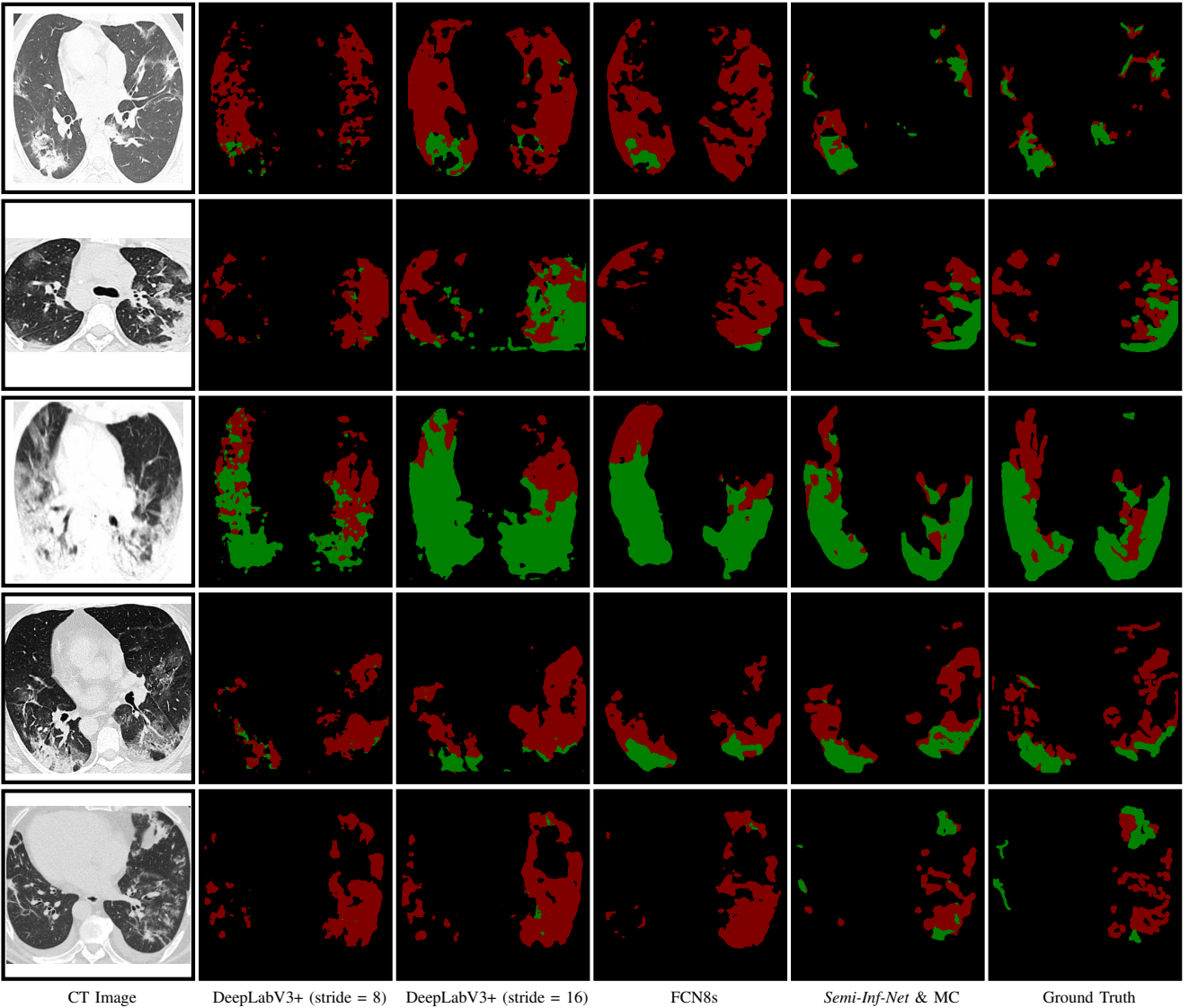

 %\label{fig:visual_results_multi}

\section{Experiments}\label{sec:Experiments}

\subsection{COVID-19 Segmentation Dataset}

As shown in Table~\ref{tab:DatasetSummary}, there is only one segmentation dataset for CT data, \ie, the COVID-19 CT Segmentation dataset~\cite{data_seg}\footnote{\url{http://medicalsegmentation.com/covid19/}},  which consists of 100 axial CT images from different COVID-19 patients. All the CT images 
were collected by the Italian Society of Medical and Interventional Radiology, and are available at here\footnote{\url{https://www.sirm.org/category/senza-categoria/covid-19}}. A radiologist segmented the CT images using different labels for identifying lung infections. 
Although this is the first open-access COVID-19 dataset for lung infection segmentation, it suffers from a small sample size, \ie, only 100 labeled images are available. 

In this work, we collected a semi-supervised COVID-19 infection segmentation dataset (\ourdataset), to leverage large-scale unlabeled CT images for augmenting the training dataset. We employ COVID-19 CT Segmentation~\cite{data_seg} as the labeled data $\mathcal{D}_{\text{Labeled}}$, which consists of 45 CT images randomly selected as training samples, 5 CT images for validation, and the remaining 50 images for testing.  The unlabeled CT images are extracted from the COVID-19 CT Collection~\cite{cohen2020covid} dataset, which consists of 20 CT volumes from different COVID-19 patients. We extracted 1,600 2D CT axial slices from the 3D volumes, removed non-lung regions, and constructed an unlabeled training dataset $\mathcal{D}_{\text{Unlabeled}}$ for effective semi-supervised segmentation.

\subsection{Experimental Settings}
\textbf{Baselines.} For the infection region experiments, we compare the proposed \ourmodel~and \textit{Semi-Inf-Net} with five classical segmentation models in the medical domain, \ie, U-Net\footnote{\url{https://github.com/MrGiovanni/UNetPlusPlus}\label{unet}}~\cite{ronneberger2015u}, U-Net++\textsuperscript{\ref{unet}}~\cite{zhou2018unetplus}, Attention-UNet\footnote{\url{https://github.com/ozan-oktay/Attention-Gated-Networks}\label{aunet}}~\cite{oktay2018attention},
Gated-UNet\textsuperscript{\ref{aunet}}~\cite{schlemper2018attention}, and Dense-UNet\footnote{\url{https://github.com/xmengli999/H-DenseUNet}}~\cite{li2018h}. For the multi-class labeling experiments, we compare our model with two cutting-edge models from the computer vision community: DeepLabV3+~\cite{chen2018encoder}, FCN8s~\cite{long2015fully} and multi-class U-Net~\cite{ronneberger2015u}.  

\textbf{Evaluation Metrics.}
Following~\cite{shan+2020lung,shi2020large}, we use three widely adopted metrics, \ie, the Dice similarity coefficient, Sensitivity (Sen.), Specificity (Spec.),  and Precision (Prec.).  We also introduce three golden metrics from the object detection field, \ie, Structure Measure~\cite{fan2017structure}, Enhance-alignment Measure~\cite{fan2018enhanced}, and Mean Absolute Error. In our evaluation, we choose $S_3$ with $Sigmoid$ function as the final prediction $S_p$. Thus, we measure the similarity/dissimilarity between final the prediction map and object-level segmentation ground-truth $G$, which can be formulated as follows:

\subsubsection{Structure Measure~($S_{\alpha}$)} This was proposed to measure the structural similarity between a prediction map and ground-truth mask, which is more consistent with the human visual system:
\begin{equation}\label{equ:s_m}
  S_\alpha = (1-\alpha)\ast S_o(S_{p}, G) + \alpha \ast S_r(S_{p}, G),
\end{equation}
where $\alpha$ is a balance factor between object-aware similarity $S_o$ and region-aware similarity $S_r$. We report $S_\alpha$ using the default setting ($\alpha=0.5$) suggested in the original paper~\cite{fan2017structure}.

\subsubsection{Enhanced-alignment Measure~($E_\phi^{mean}$)} 
This is a recently proposed metric for evaluating both local and global similarity between two binary maps. The formulation is as follows:
\begin{equation}\label{equ:e_m}
  E_{\phi} = \frac{1}{w \times h} \sum_{x}^{w} \sum_{y}^{h} \phi(S_{p}(x, y), G(x, y)),
\end{equation}
where $w$ and $h$ are the width and height of ground-truth $G$, and $(x,y)$ denotes the coordinate of each pixel in $G$. Symbol $\phi$ is the enhanced alignment matrix. We obtain a set of $E_{\phi}$ by converting the prediction $S_p$ into a binary mask with a threshold from 0 to 255. In our experiments, we report the mean of $E_{\xi}$ computed from all the thresholds.

\subsubsection{Mean Absolute Error~(MAE)} This measures the pixel-wise error between $S_p$ and $G$, which is defined as:
\begin{equation}\label{equ:mae}
  \textit{MAE} = \frac{1}{w \times h} \sum_{x}^{w} \sum_{y}^{h} |S_p(x,y) - G(x,y)|.
\end{equation}

\subsection{Segmentation Results}
% In this section, we present the quantitative and qualitative results to validate the performance of the proposed \ourmodel. 

\subsubsection{Quantitative Results}
To compare the infection segmentation performance, we consider the two state-of-the-art models U-Net and U-Net++.
Quantitative results are shown in \tabref{tab:InfectionRegions}. As can be seen, the proposed \ourmodel~outperforms U-Net and U-Net++ in terms of Dice, $S_\alpha$, $E_\phi^{mean}$, and MAE by a large margin. We attribute this improvement to our implicit reverse attention and explicit edge-attention modeling, which provide robust feature representations. In addition, by introducing the semi-supervised learning strategy into our framework, we can further boost the performance with a 5.7\% improvement in terms of Dice.
% Our~\ourmodel~also outperforms the state-of-the-art results reported in the literature. For instance, as clarified in \cite{COVID-19-SegBenchmark}, the infection segmentation Dice score given by nnU-Net \cite{Fabian2020Automated} is 0.673, which is lower than the score (0.739) given by our method.

\begin{table*}[!t]
  \centering
  \footnotesize
  \renewcommand{\arraystretch}{1.1}
  \setlength\tabcolsep{9.0pt}
  \caption{\small Quantitative results of infection regions on our   \ourdataset~dataset. 
  %The best two results are shown in \tr{red} and \tb{blue} fonts.
  }\label{tab:InfectionRegions} 
  \begin{tabular}{l|lrr||c|c|c|c|c|c}
  \hline
  \rowcolor{mygray}
   Methods & Backbone & Param. & FLOPs & Dice & Sen. & Spec. & $S_{\alpha}$ & $E_\phi^{mean}$ & MAE \\
  \hline
   U-Net~\cite{ronneberger2015u} 
  & VGG16 & 7.853 M & 38.116 G & 0.439 & 0.534 & 0.858 & 0.622 & 0.625 & 0.186\\
  
   Attention-UNet~\cite{oktay2018attention}
  & VGG16 & 8.727 M & 31.730 G & 0.583 & 0.637 & 0.921 & 0.744 & 0.739 & 0.112\\
  
   Gated-UNet~\cite{schlemper2018attention} 
  & VGG16 & 175.093 K  & 714.419 M & 0.623& 0.658 & 0.926 & 0.725 & 0.814 & 0.102\\
  
   Dense-UNet~\cite{li2018h} 
  & DenseNet161 & 45.082 M  & 43.785 G & 0.515 & 0.594 & 0.840 & 0.655 & 0.662 & 0.184 \\
  
   U-Net++~\cite{zhou2018unetplus} 
  & VGG16 & 9.163 M & 65.938 G & 0.581  & 0.672 &  0.902  & 0.722 & 0.720 & 0.120\\
  
  %& nnU-Net~\cite{Fabian2020Automated} 
  %& xxx & xxx & xxx & xxx & xxx & xxx\\
  \hline
  \hline
   \textbf{\ourmodel~(Ours)} 
  & Res2Net~\cite{gao2019res2net} & 33.122 M & 13.922 G & \textbf{\tb{0.682}} & \textbf{\tb{0.692}} & \textbf{\tb{0.943}} &  \textbf{\tb{0.781}} & \textbf{\tb{0.838}} & \textbf{\tb{0.082}}\\
   \textbf{\textit{Semi-}\ourmodel~(Ours)}
  & Res2Net~\cite{gao2019res2net} & 33.122 M & 13.922 G & \textbf{\tr{0.739}} & \textbf{\tr{0.725}} & \textbf{\tr{0.960}} &  \textbf{\tr{0.800}} & \textbf{\tr{0.894}} & \textbf{\tr{0.064}}\\
  \hline
  \end{tabular}
\end{table*}

% \begin{table}[t!]
%   \centering
%   \footnotesize
%   \renewcommand{\arraystretch}{1.35}
%   \setlength\tabcolsep{2.8pt}
%   \caption{\small Quantitative results of infection regions on our   \ourdataset~dataset. The best two results are shown in \tr{red} and \tb{blue} fonts.
%   }\label{tab:InfectionRegions}
%   \begin{tabular}{ll||cccccccc}
%   \hline
%   % \multicolumn{1}{l|}{Methods} &\multicolumn{3}{c|}{\emph{MCL}}\\
%   \rowcolor{mygray}
%   &Methods & Dice & Sen. & Spec. & $S_{\alpha}$ & $E_\phi^{mean}$ & MAE \\
%   \hline
% %   & \multicolumn{1}{l}{Infection Regions}\\
% %   \hline
%   %\multirow{5}{*}{\begin{sideways}Infection Regions\end{sideways}} & 
%   & U-Net (MICCAI'15)~\cite{ronneberger2015u} & 0.439 & 0.534 & 0.858 & 0.622 & 0.625 & 0.186\\
%   &U-Net++ (TMI'19)~\cite{zhou2018unetplus} & 0.581  & 0.672 &  0.902  & 0.722 & 0.720 & 0.120\\
%   %& PraNet~\cite{fan2020PraNet} & 0.702 & 0.808 & 0.948 & 0.761 & 0.812\\
%   \hline
%   \hline
%   & \textbf{\ourmodel~(Ours)} & \textbf{\tb{0.682}} & \textbf{\tb{0.692}} & \textbf{\tb{0.943}} &  \textbf{\tb{0.781}} & \textbf{\tb{0.838}} & \textbf{\tb{0.082}}\\
%   & \textbf{\textit{Semi-}\ourmodel~(Ours)} & 
%   \textbf{\tr{0.739}} & \textbf{\tr{0.725}} & \textbf{\tr{0.960}} &  \textbf{\tr{0.800}} & \textbf{\tr{0.894}} & \textbf{\tr{0.064}} \\
%   \hline
%   \end{tabular}
% \end{table}

As an assistant diagnostic tool, the model is expected to provide more detailed information regarding the infected areas. Therefore, we extent to our model to the multi-class (\ie, GGO and consolidation segmentation) labeling.
\tabref{tab:Ground-glass} shows the quantitative evaluation on our \ourdataset~dataset, where ``\textit{Semi-}\ourmodel~\& FCN8s" and  ``\textit{Semi-}\ourmodel~\& MC" denote the combinations of our \textit{Semi-}\ourmodel with FCN8s~\cite{long2015fully} and multi-class U-Net~\cite{ronneberger2015u}, respectively.
Our ``\textit{Semi-}\ourmodel~\& MC" pipeline achieves the competitive performance on GGO segmentation in most evaluation metrics. 
For more challenging consolidation segmentation, the proposed pipeline also achieves best results.
For instance, in terms of Dice, our method outperforms the cutting-edge model, Multi-class U-Net~\cite{ronneberger2015u}, by 12\% on average segmentation result.
Overall, the proposed pipeline performs better than existing state-of-the-art models on multi-class labeling on consolidation segmentation and average segmentation result in terms of Dice and $S_\alpha$.

\begin{table*}[thp!]
  \centering
  \footnotesize
  \renewcommand{\arraystretch}{1.1}
  \setlength\tabcolsep{1.2pt}
  \caption{\small Quantitative results of ground-glass opacities and consolidation on our \ourdataset~dataset. The best two results are shown in \tr{red} and \tb{blue} fonts. Please refer to our manuscript for the complete evaluations.
  }\label{tab:Ground-glass}
  \begin{tabular}{ll|cccccc||cccccc||cccccc}
  \hline
   &  & \multicolumn{6}{c||}{\tabincell{c}{Ground-Glass Opacity}} &\multicolumn{6}{c||}{\tabincell{c}{Consolidation}}
   &\multicolumn{6}{c}{\tabincell{c}{Average}} \\
   \cline{3-20}
   & Methods & Dice & Sen. & Spec. & $S_\alpha$ & $E_\phi^{mean}$ & MAE & Dice & Sen. & Spec. & $S_\alpha$  & $E_\phi^{mean}$ & MAE & Dice & Sen. & Spec. & $S_\alpha$ & $E_\phi^{mean}$ & MAE\\
  \hline
     & DeepLabV3+ (stride=8)~\cite{chen2018encoder} & 0.375 & 0.478 & 0.863 & 0.544 & 0.675 & 0.123 & 0.148 & 0.152 & 0.738 & 0.500 & 0.523 & 0.064 & 0.262 & 0.315 & 0.801 & 0.522 & 0.599 & 0.094 \\
     
     & DeepLabV3+ (stride=16)~\cite{chen2018encoder} & 0.443 & \textbf{\tb{0.713}} & 0.823 & 0.548 & 0.655 & 0.156 & 0.238 & 0.310 & 0.708 & 0.517 & \textbf{\tb{0.606}} & 0.077 & 0.341 & \textbf{\tb{0.512}} & 0.766 & 0.533 & 0.631 & 0.117 \\
     
     & FCN8s~\cite{long2015fully}& 0.471 & 0.537 & 0.905 & 0.582 & 0.774 & 0.101 & 0.279 & 0.268 & 0.716 & 0.560 & 0.560 & 0.050 & 0.375 & 0.403 & 0.811 & 0.571 & 0.667 & 0.076\\
     
     & Multi-class U-Net & 0.441 & 0.343 & \textbf{\tr{0.984}} & 0.588 & 0.714 & 0.082 & \textbf{\tb{0.403}} & \textbf{\tb{0.414}} & \textbf{\tr{0.967}} & \textbf{\tb{0.577}} & \textbf{\tr{0.767}} & 0.055 & 0.422 & 0.379 & \textbf{\tr{0.976}} & 0.583 & \textbf{\tb{0.741}} & 0.066 \\
     \hline
     \hline
     & \textbf{\textit{Semi-}\ourmodel~\& FCN8s} &  \textbf{\tr{0.646}} & \textbf{\tr{0.720}} & 0.941 & \textbf{\tr{0.711}} & \textbf{\tb{0.882}} & \textbf{\tb{0.071}}  & 0.301 & 0.235 & \textbf{\tb{0.808}} & 0.571 & 0.571  & \textbf{\tr{0.045}} & \textbf{\tb{0.474}} & 0.478 & 0.875 & \textbf{\tb{0.641}} & 0.723 &  \textbf{\tb{0.058}}\\
      & \textbf{\textit{Semi-}\ourmodel~\& MC} &  \textbf{\tb{0.624}} & 0.618 & \textbf{\tb{0.966}} & \textbf{\tb{0.706}} & \textbf{\tr{0.889}} & \textbf{\tr{0.067}} & \textbf{\tr{0.458}} & \textbf{\tr{0.509}} & \textbf{\tr{0.967}} & \textbf{\tr{0.603}} & \textbf{\tr{0.767}} & \textbf{\tb{0.047}} & \textbf{\tr{0.541}} & \textbf{\tr{0.564}} & \textbf{\tb{0.967}} & \textbf{\tr{0.655}} & \textbf{\tr{0.828}} & \textbf{\tr{0.057}} \\
  \hline
  \end{tabular}
\end{table*}

\subsubsection{Qualitative Results} 
The lung infection segmentation results, shown in Fig.~\ref{fig:visual_results_single}, indicate that our \textit{Semi-}\ourmodel~and \ourmodel~outperform the baseline methods remarkably. Specifically, they yield segmentation results that are close to the ground truth with much less mis-segmented tissue. In contrast, U-Net gives unsatisfactory results, where a large number of mis-segmented tissues exist. U-Net++ improves the results, but the performance is still not promising. The success of \ourmodel~is owed to our coarse-to-fine segmentation strategy, where a parallel partial decoder first roughly locates lung infection regions and then multiple edge attention modules are employed for fine segmentation. This strategy mimics how real clinicians segment lung infection regions from CT slices, and therefore achieves promising performance. In addition, the advantage of our semi-supervised learning strategy is also confirmed by Fig.~\ref{fig:visual_results_single}. As can be observed, compared with \ourmodel, \textit{Semi-}\ourmodel~yields segmentation results with more accurate boundaries. In contrast, \ourmodel~gives relatively fuzzy boundaries, especially in the subtle infection regions.

We also show the multi-class infection labeling results in Fig.~\ref{fig:visual_results_multi}. As can be observed, our model,  \textit{Semi-}\ourmodel~\& MC, consistently performs the best among all methods.
It is worth noting that both GGO and consolidation infections are accurately segmented by \textit{Semi-}\ourmodel~\& MC, which further demonstrates the advantage of our model. In contrast, the baseline methods, DeepLabV3+ with different strides and FCNs, all obtain unsatisfactory results, where neither GGO and consolidation infections can be accurately segmented.

\subsection{Ablation Study}
In this subsection, we conduct several experiments to validate the performance of each key component of our \textit{Semi-}\ourmodel, including the PPD, RA, and EA modules. 

%including parallel Partial Decoder (PPD), Reverse Attention (RA), and Edge Attention (EA). 
%To largely alleviate the issues introduced by limited training data, we further validate the effectiveness of the proposed semi-supervised learning strategy.

\subsubsection{Effectiveness of PPD} To explore the contribution of the parallel partial decoder, we derive two baselines: No.1 (backbone only) \& No.3 (backbone+PPD) in \tabref{tab:ablation}. The results clearly show that PPD is necessary for boosting performance. 

\subsubsection{Effectiveness of RA} We investigate the importance of the RA module. From \tabref{tab:ablation}, we observe that No.4 (backbone + RA) increases the backbone performance (No.1) in terms of major metrics, \eg, Dice, Sensitivity, MAE, \etc. This suggests that introducing the RA component can enable our model to accurately distinguish true infected areas.

\subsubsection{Effectiveness of PPD \& RA} We also investigate the importance of the combination of the PPD and RA components (No.6). As shown in \tabref{tab:ablation}, No.4 performs better than other settings (\ie, No.1$\sim$No.4) in most metrics. These improvements demonstrate that the reverse attention together with the parallel partial decoder are the two 
central components responsible for the good performance of 
\ourmodel.

\subsubsection{Effectiveness of EA} Finally, we investigate the importance of the EA module. From these results in \tabref{tab:ablation} (No.2 vs. No.1, No.5 vs. No.4, No.7 vs. No.6), it can be clearly observed that EA module effectively improves the segmentation performance in our \ourmodel.
 
\begin{table}[t!]
  \centering
  \footnotesize
  \renewcommand{\arraystretch}{1.1}
  \setlength\tabcolsep{1.3pt}
  \caption{\small Ablation studies of our \textit{Semi-}\ourmodel. The best two results are shown in \tr{red} and \tb{blue} fonts. 
  }\label{tab:ablation}
  \begin{tabular}{ll||cccccccc}
  \hline
  % \multicolumn{1}{l|}{Methods} &\multicolumn{3}{c|}{\emph{MCL}}\\
  \rowcolor{mygray}
   &Methods & Dice & Sen. & Spec. &$S_{\alpha}$ &$E_\phi^{mean}$& MAE \\
  \hline
  \hline
  %\multirow{5}{*}{\begin{sideways}Infection Regions\end{sideways}} & 
   &(No.1) Backbone & 0.442 & 0.570 & 0.825 & 0.651 & 0.569 & 0.207\\
   &(No.2) Backbone+EA & 0.541 & 0.665 & 0.807 & 0.673 & 0.659 & 0.205 \\
   &(No.3) Backbone+PPD & 0.669 & 0.744 & 0.880 & 0.720 & 0.810 & 0.125 \\
   &(No.4) Backbone+RA& 0.625 & \textbf{\tr{0.826}} & 0.809 & 0.668 & 0.736 & 0.177\\
   &(No.5) Backbone+RA+EA & \textbf{\tb{0.672}} & \textbf{\tb{0.754}} & 0.882 & 0.738 & 0.804 & 0.122\\
   &(No.6) Backbone+PPD+RA& 0.655 & 0.690 & \textbf{\tb{0.927}} & \textbf{\tb{0.761}} & \textbf{\tb{0.812}} & \textbf{\tb{0.098}}\\
  %& (No.5) Backbone+PPD+RA+EA & \textbf{\tr{0.753}} & 0.782 & \textbf{\tb{0.962}} & \textbf{\tr{0.792}} & \textbf{\tr{0.894}} &\textbf{\tr{0.067}} \\
  \hline
    & (No.7) Backbone+PPD+RA+EA & \textbf{\tr{0.739}} & 0.725 & \textbf{\tr{0.960}} &  \textbf{\tr{0.800}} & \textbf{\tr{0.894}} & \textbf{\tr{0.064}} \\
  \hline
  \end{tabular}
\end{table}

\subsection{Evaluation on Real CT Volumes}
In the real application, each CT volume has multiple slices, where most slices could have no infections. To further validate the effectiveness of the proposed method on real CT volume, we utilized the recently released COVID-19 infection segmentation dataset~\cite{data_seg}, which consists of 638 slices (285 non-infected slices and 353 infected slices) extracting from 9 CT volumes of real COVID-19 patients as test set for evaluating our model performance. The results are shown in  Tables~\ref{tab:RealData-all}. Despite containing non-infected slices, our method still obtains the best performance. Because  we employed two datasets for semi-supervised learning, \ie, labeled data with 100 infected slices (50 training, 50 testing), and unlabeled data with 1600 CT slices from real volumes. The unlabeled data contains a lot of non-infected slices to guarantee our model could deal with non-infected slices well.   Moreover, our \ourmodel~is a general infection segmentation framework, which could be easily implemented for other type of infection.

\begin{table}[!t]
	\centering
	\footnotesize
	\renewcommand{\arraystretch}{1.1}
	\setlength\tabcolsep{6.0pt}
	\caption{\small Performances  on nine \textit{real CT volumes} with 638 slices (285 non-infected and 353 infected slices). The best two results are shown in \tr{red} and \tb{blue} fonts. 
	}\label{tab:RealData-all}
	\begin{tabular}{l||ccccc}
	\hline
	\rowcolor{mygray}
		  Model                                              &     Dice     &     Sen.     &    Spec.     & Prec. &     MAE      \\
	\hline
		  U-Net~\cite{ronneberger2015u}  & 0.308 & 0.678 & 0.836 & 0.265 & 0.214 \\
		  Attention-UNet~\cite{oktay2018attention} &    0.466 & 0.723 & 0.930 & 0.390 & 0.095 \\
		  Gated-UNet~\cite{schlemper2018attention}  & 0.447 & 0.674 & 0.956 & 0.375 & 0.066 \\
		  Dense-UNet~\cite{li2018h}                 & 0.410 & 0.607 & \textbf{\tr{0.977}} & 0.415 & 0.167 \\
		  U-Net++~\cite{zhou2018unetplus}           & 0.444 & \textbf{\tr{0.877}} & 0.929 & 0.369 & 0.106 \\ \hline\hline
		  \textbf{\ourmodel~(Ours)}                          & \textbf{\tb{0.579}} & \textbf{\tb{0.870}} & \textbf{\tb{0.974}} & \textbf{\tb{0.500}} & \textbf{\tb{0.047}} \\
		  \textbf{\textit{Semi}-\ourmodel~(Ours)}                     & \textbf{\tr{0.597}} & 0.865 & \textbf{\tr{0.977}} & \textbf{\tr{0.515}} & \textbf{\tr{0.033}} \\ \hline
	\end{tabular}
\end{table}

\subsection{Limitations and Future Work}

 Although the our \ourmodel~achieved promising results in segmenting infected regions, there are some limitations in the current model. First, the \ourmodel~focuses on lung infection segmentation for COVID-19 patients. However, in clinical practice, it often requires to classify COVID-19 patients and then segment the infection regions for further treatment. Thus, we will study an AI automatic diagnosis system, which integrates COVID-19 detection, lung infection segmentation, and infection regions quantification into a unified framework. Second, for our multi-class infection labeling framework, we first apply the \ourmodel~to obtain the infection regions, which can be used to guide the multi-class labeling of different types of lung infections. It can be seen that we conduct a two-step strategy to achieve multi-class infection labeling, which could lead to sub-optimal learning performance. In future work, we will study to construct an end-to-end framework to achieve this task. Besides, due to the limited size of dataset, we will use the Generative Adversarial Network (GAN) \cite{zhou2020hi} or Conditional Variational Autoencoders (CVAE)~\cite{Zhang2020UCNet} to synthesize more samples, which can be regarded as a form of data augmentation to enhance the segmentation performance. 
 Moreover, our method may have a bit drop in accuracy when considering non-infected slices. Running a additional slice-wise classifier (\eg, infected vs non-infected) for selecting the infected slice is an effective solution for avoiding the performance drop on non-infected slices. 

\section{Conclusion}

In this paper, we have proposed a novel COVID-19 lung CT infection segmentation network, named \ourmodel, which utilizes an implicit reverse attention and explicit edge-attention to improve the identification of infected regions. 
Moreover, we have also provided a semi-supervised solution, \textit{Semi-}\ourmodel, to alleviate the shortage of high quality labeled data. Extensive experiments on our \ourdataset~dataset and real CT volumes have demonstrated that the proposed \ourmodel~and \textit{Semi-}\ourmodel~outperform the cutting-edge segmentation models and advance the state-of-the-art performances. Our  system has great potential to be applied in assessing the diagnosis of COVID-19, \eg, quantifying the infected regions, monitoring the longitudinal disease changes, and  mass screening processing.  
Note that the proposed model is able to detect the objects with 
low intensity contrast between infections and normal tissues. 
This phenomenon is often occurs in nature camouflage objects. In the future, we plan to apply our \ourmodel~to other related tasks, such as 
polyp segmentation~\cite{fang20PraNet}, product defects detection, camouflaged animal detection~\cite{fan2020Camouflage}. Our code and dataset have been released at: \href{https://github.com/DengPingFan/Inf-Net}{https://github.com/DengPingFan/Inf-Net}

{
\bibliographystyle{IEEEtran}
\bibliography{covid}
}
 
\end{document}